\documentclass[%
 reprint,
superscriptaddress,
 amsmath,amssymb,
 aps,
]{revtex4-2}

\usepackage{graphicx}% Include figure files
\usepackage{dcolumn}% Align table columns on decimal point
\usepackage{bm}% bold math
\usepackage{algorithm2e}
\usepackage{hyperref}
\RestyleAlgo{ruled}
\usepackage{xcolor}
\usepackage[normalem]{ulem}

\begin{document}

\preprint{APS/123-QED}

\title{Advancing Algorithm to Scale and Accurately Solve Quantum Poisson Equation on Near-term Quantum Hardware
}

%\author{\IEEEauthorblockN{1\textsuperscript{st} Kamal K. Saha}
%\IEEEauthorblockA{\textit{Center For Research Computing} \\
%\textit{University Of Notre Dame}\\
%Notre Dame, Indiana \\
%ksaha@nd.edu}
%\and
%\IEEEauthorblockN{2\textsuperscript{nd} Walter Robson}
%\IEEEauthorblockA{\textit{Center for Research Computing} \\
%\textit{University Of Notre Dame}\\
%Notre Dame, Indiana \\
%wrobson@nd.edu}
%\and
%\IEEEauthorblockN{3\textsuperscript{rd} Connor Howington}
%\IEEEauthorblockA{\textit{Center For Research Computing} \\
%\textit{University Of Notre Dame}\\
%Notre Dame, Indiana \\
%chowingt@nd.edu}

\author{Kamal K. Saha}
\email[Corresponding author: ]{ksaha@nd.edu}
\affiliation{Center for Research Computing, University of Notre Dame, Notre Dame, IN 46556, USA}

\author{Walter Robson}
% \email{wrobson@nd.edu}
\affiliation{Department of Computer Science and Engineering, University of Notre Dame, Notre Dame, IN 46556, USA}

\author{Connor Howington}
\affiliation{Center for Research Computing, University of Notre Dame, Notre Dame, IN 46556, USA}
% \email{chowingt@nd.edu}

\author{In-Saeng Suh}
%\author[0000-0002-6923-6455]{In-Saeng Suh}
% \email[]{suhi@ornl.gov}
\affiliation{National Center for Computational Sciences, Oak Ridge National Laboratory, Oak Ridge, TN 37830, USA}
\affiliation{Center for Research Computing, University of Notre Dame, Notre Dame, IN 46556, USA}

\author{Zhimin Wang}
\affiliation{Faculty of Information Science and Engineering, Ocean University of China, Qingdao 266100, China}

\author{Jaroslaw Nabrzyski}
% \email{naber@nd.edu}
\affiliation{Center for Research Computing, University of Notre Dame, Notre Dame, IN 46556, USA}

%}

\date{\today}

\begin{abstract}
The Poisson equation has many applications across the broad areas of science and engineering. Most quantum algorithms for the Poisson solver presented so far either suffer from lack of accuracy and/or are limited to very small sizes of the problem, and thus have no practical usage. Here we present an advanced quantum algorithm for solving the Poisson equation with high accuracy and dynamically tunable problem size. After converting the Poisson equation to a linear system through the finite difference method, we adopt the HHL algorithm as the basic framework. Particularly, in this work we present an advanced circuit that ensures the accuracy of the solution by implementing non-truncated eigenvalues through eigenvalue amplification, as well as by increasing the accuracy of the controlled rotation angular coefficients, which are the critical factors in the HHL algorithm. Consequently, we are able to drastically reduce the relative error in the solution while achieving higher success probability as the amplification level is increased. We show that our algorithm not only increases the accuracy of the solutions but also composes more practical and scalable circuits by dynamically controlling problem size in NISQ devices. We present both simulated and experimental results and discuss the sources of errors. Finally, we conclude that though overall results on the existing NISQ hardware are dominated by the error in the \textsl{CNOT} gates, this work opens a path to realizing a multidimensional Poisson solver on near-term quantum hardware.
\end{abstract}

\maketitle

%\begin{IEEEkeywords}
%Quantum Computing, Poisson Equation
%\end{IEEEkeywords}

\section{Introduction}\label{sec.I}
The Poisson equation is a second-order partial differential equation widely used in various fields of science and engineering.
In general, in order to solve the Poisson equation numerically, projection methods such as collocation, spectral, and boundary element methods as well as finite-difference methods \cite{Poisson} are used.
The core of these methods is to approximate the solution of the Poisson equation as the solution of a linear system. However, since the dimension of the linear system obtained from the discrete Poisson equation is generally very large, solving such a system demands much computational time. Therefore, the Poisson equation is a problem well suited to quantum computing, a faster and more powerful computation paradigm \cite{Feynman1982} than classical computing.

A series of quantum algorithms \cite{Leyton2008, Berry2014, Berry2017, Childs2020, Childs2021, Costa2019, Arrazola2019, Dervovic2018, HHL2009, Cao2012, Childs2017, Berry2015, Kalaj2019, Huang2021, Subasi2019, Liu2021, Sato2021, Saito2021, Cao2013, Wang2020a} have been developed to solve linear equation systems, which have shown significant speedups over their classical counterparts. Recently, variational quantum algorithms (VQAs) \cite{McClean2016, Cerezo2021, Zhou2020, Barron2020}, which have already shown some promise for use on so-called noisy intermediate-scale quantum (NISQ) devices \cite{Preskill2018} are being adopted to solve the Poisson equation \cite{Liu2021,Sato2021}. 
% However, even though the VQAs method can solve the one-dimensional Poisson equation under limited conditions, it still can not be extended to a general framework for solving higher dimensional Poisson equations. 
From the experimental point of view, while VQAs-based approaches have some advantages, such as generally using relatively shallow quantum circuits or requiring fewer quantum measurements, they still have challenges in optimizing a set of parameters, especially on larger problems \cite{Liu2021}. In addition, instead of producing the direct solution of the Poisson equation, these methods rely on an expectation of certain observables limiting them to be coupled with other general problems, and thus may have a limited use case. An improved iterative method for the HHL algorithm \cite{HHL2009} has been proposed to solve linear system of equations in Ref. \cite{Saito2021}. Even though they obtained a more accurate solution by increasing the number of iterations with the same number of measurements, they still have challenges in improving the error convergence speed compared to the state vector calculations.

However, in the context of the quantum circuit model, Cao et al. \cite{Cao2013} first used the original HHL algorithm \cite{HHL2009} to solve the Poisson equation. Later, our co-author Wang et al. \cite{Wang2020a} pointed out a bottleneck of Cao's algorithm where the controlled rotation is implemented by the arc sine function evaluation. In other words, the bottleneck comes from the process of performing a linear mapping from state  $|\lambda_{j}\rangle$ to $\lambda_{j}^{-1}|\lambda_j \rangle$, where $\lambda_{j}$ represents the eigenvalues of a matrix of the linear system of equations. More precisely, after having the eigenvalue state $|\lambda_{j}\rangle$ by phase estimation, Cao et al. evaluate the reciprocal state $|1/\lambda_{j}\rangle$ through the Newton iteration method. After that, the binary state of $|1/\lambda_{j}\rangle$ is converted to the probability amplitude $1/\lambda_j$ through the controlled $R_y$ rotations with the angle of $\theta = {\rm arcsin}(1/\lambda_j)$, where the arc sine function is evaluated by the cut-and-try method. Since the cost of calculating the sine function is $O(m^3)$ where $m$ is the number of qubits of input register, then the evaluation cost of the arc sine function is $O(m^4)$ \cite{Wang2020a}.

Wang's approach resolved the bottleneck of Cao's algorithm and developed a quantum fast Poisson solver with complete and modular circuit representation. First, they proposed a new way of implementing the controlled rotation in the HHL algorithm. That is, they introduced a method in which they take the state $|\lambda_{j}\rangle$ to $\lambda_{j}^{-1}|\lambda_j \rangle$ directly without passing through the $|1 / \lambda_{j} \rangle$ state. In this process, they adopted a novel method called qFBE (quantum function-value binary expansion) to evaluate the arc cotangent function \cite{Borwein1995, Wang2020b}. With this method, they reduced the cost of the problem from $O(m^4)$ to $O(m^3)$. Second, they developed the inverse qFBE method to compute the cosine function in order to simplify the Hamiltonian simulation subroutine of HHL, making the circuit design easier and more modular. Finally, they also exploited quantum algorithms for solving the reciprocal and square root operations using the classical non-restoring method \cite{Sutikno2011}. By developing a new way of implementing the controlled rotation within HHL and quantum circuits for solving the Poisson equation, they not only reduced the algorithm's complexity but also made the circuit complete and implementable. However, in reducing the cost and complexity of the quantum circuit, Wang et al. truncated both the eigenvalues of the matrix and the rotation angular coefficients. As a result, numerical errors are accumulated, and eventually that compromises the accuracy of the solution of the Poisson equation. 

Even though these past works, including Cao's and Wang's methods, improved the quantum algorithm and circuit for the Poisson solver, they still either suffered from lack of accuracy and/or were limited to demonstrating only a very small size of the problem, and thus their practical usage is limited. Some of these works focus on minimizing the error in their approaches or in the overall solutions without directly presenting the actual or direct solution of the Poisson equation \cite{Liu2021,Sato2021,Saito2021}, or some others appeared to suggest the feasibility of their methods on quantum hardware without even clearly discussing or validating their works on any hardware \cite{Liu2021,Sato2021}.

% -- Variational method limited a set of parameter which are difficult to optimize for a larger problem
% -- They don't have any experimental validation or discussion for such problems. 

This paper advances the algorithm for solving the Poisson equation in several aspects: (1) Improve the precision of phase estimation by increasing the accuracy of the eigenvalues. Unlike the previous approach \cite{Wang2020a} where only the integer part of the eigenvalues was encoded, we implement non-truncated eigenvalues through eigenvalue amplification. We will see that this has a clear impact in drastically reducing the error in the solution of the Poisson solver; (2) The rotation angles are calculated with full accuracy, which is also essential for ensuring the overall accuracy of the solution; (3) Without compromising any accuracy of the algorithm, during the run-time, our implementation uses an optimized number of qubits representing the rotation angles. We also optimize the \textsl{CNOT} gates usage, which is one of the primary sources of error in an experiment; (4) Solutions of the Poisson equations with larger problem size to $7\times 7$ and $15\times 15$ are demonstrated. In fact, our implementation with dynamic allocation of qubits in different segments of the algorithm ensures easy adaptation of this method for solving real-world problems; (5) The possibilities and difficulties of implementing the algorithm on the real quantum hardware are discussed for the first time. This also includes experimenting with the circuit mapping, error mitigation, etc. on the NISQ devices and presenting a vision for near-term hardware; (6) Finally, the algorithm is implemented using Qiskit package \cite{Qiskit}, which would bring advantages for practical use. We believe all these aspects are necessary to advance the study of quantum Poisson solvers.

We also want to make it clear that in this work our main focus is  advancing the hybrid algorithm to accurately simulate the Poisson equation with realistic problem sizes, while also exploring the experimental feasibility of those problems. In particular, we aim to push the scalability of our proposed Poisson solver to larger practical problems on both simulators and real quantum devices. While testing these problems, we also identify the key limiting factors against applying the algorithm to large problems and implement some optimization methods in terms of the number of qubits and gates. We explain that with the current state of the technology, it is difficult to realize a complete quantum description of the algorithm due to its high resource costs. However, we discuss pathways to further improve this hybrid approach in both simulation and experimental environments.

For demonstrating the circuit, we present both simulated and experimental results, discuss the sources of errors, and eliminate them. The Matrix Product State (MPS) simulator is used for simulation, and the experiment is done on IBM's \textsl{ibmq{\textunderscore}manila} and \textsl{ibmq{\textunderscore}brooklyn} quantum backends \cite{IBMQ}. We examine the measurement error mitigation on a small system and also discuss how the overall results of the Poisson equation on the currently available quantum hardware are dominated by the error in the \textsl{CNOT} gates.

We have extended the existing algorithm from Wang et al. beyond a single proof of concept to a fully dynamic, scalable body of work that can be used for numerous applications in mixed computing algorithms. Wang’s QRUNES \cite{Qrunes}-based machine instructions have been abstracted to more usable Qiskit functions, allowing us to perform fine-tuning of the different register sizes so that we can identify key areas of inaccuracy and compare qubit tradeoffs. This is important because the primary limiting factor in accuracy is the total number of qubits in the circuit, and qubits are at a premium in NISQ hardware. Our code is readable and easily usable, allowing a true “black box” approach to be taken to solving the most computationally intensive part of Poisson applications.

The paper is organized as follows. In Sec.\,II, we adopt the finite difference method to discretize the Poisson equation to obtain a linear system. In Sec.\,III, we describe the quantum algorithm and circuit design for each module and our algorithm in detail. In Sec.\,IV, we explain the algorithm improvement and circuit optimization. We show simulated results of different sizes of problems and their improvements, and we discuss more about algorithm scaling and success probability in Sec.\,V. In Sec.\,VI, we demonstrate our improved quantum circuit for the Poisson solver on IBM quantum hardware and discuss error mitigation. Finally, we conclude our works in Sec.\,VII. 

\section{Overview of the Problem}\label{overview}
The goal of this work is to implement an efficient quantum algorithm solving the multi-dimensional Poisson equation with boundary conditions. Let us consider the Poisson equation defined in an open bounded domain $\Omega \subset \Re^d$, where $d$ is the number of spatial dimensions.
\begin{align}
	& -\nabla^2 v(x) = b(x), \ x \ \text{in} \ \Omega \ \
    \\
	& v(x) = 0,  \ x \ \text{on} \ \delta\Omega \ \ \Omega = (0,1)^d
\end{align}
where $\delta \Omega$ is the boundary of $\Omega$ and $b(x)$ is a given smooth function representing different problem applications, such as charge or velocity distribution. One way to solve this problem is to discretize $\Omega$ to $N^\prime = N + 1$ grid points in each dimension, where $N$ is an exponent of base $2$. The solution $v(x)$ is a vector of $(N-1)^d$ entries. 

In this work, we focus on the one-dimensional Poisson equation with Dirichlet boundary conditions. Using the central-difference approximation 
to discretize the second-order derivative, Eq. (1) can be converted to finite difference form as
\begin{equation}
	A \cdot
    \begin{pmatrix}
    v_1 \\ 
    v_2 \\ 
    \vdots \\ 
    v_{N-1}
    \end{pmatrix} 
	 = \frac{1}{h^{2}}
    \begin{pmatrix}
     2      & -1      &        &  0     \\
    -1      & \ddots  & \ddots &        \\
            & \ddots  & \ddots & -1     \\
     0      &         & -1     &  2 
    \end{pmatrix}
    \cdot
    \begin{pmatrix}
    v_1 \\ v_2 \\ \vdots \\ v_{N-1}
    \end{pmatrix}    
	= \begin{pmatrix}
    b_1 \\ b_2 \\ \vdots \\ b_{N-1}
    \end{pmatrix}
    \label{eq.3}
\end{equation}
We now have the $N-1$ linear equation system, i.e., $A |v\rangle = |b\rangle$ to be solved. Here $A$ is a Hermitian matrix with dimensions of $(N-1) \times (N-1)$, and the mesh size $h$ equals $1/N$. 
The eigenvalues of $A$ are $\lambda_j = 4N^2 {\rm sin}^2(j \pi / 2N)$, and its corresponding eigenvectors are $u_j (k) = \sqrt{2/N} {\rm sin}(j \pi k /N)$ \cite{Demmel1997}.

\begin{figure}[t]
    \centering
    \includegraphics[scale=1.15]{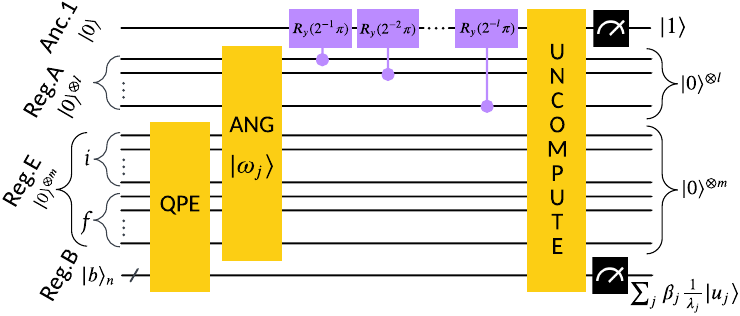}
	\caption{The overall circuit representation of the algorithm for solving the one-dimensional Poisson equation. The numbers of qubits of registers A, E, and B are $l$, $m$, and $n$, respectively. Here $m=i+f$, where $i$ and $f$ number of qubits in reg.\,E hold the integer and fractional parts of the eigenvalue. $|\omega_j\rangle$ is the angular coefficient evolved from the approximated eigenvalue $|\lambda_j\rangle$, the output of the QPE. The input  
	$|b\rangle_n = \sum_{i=1}^{2^n-1} b_i|i\rangle$ is prepared and stored in register B.}
\label{fig.1}
\end{figure}

The best classical algorithms for solving this problem run polynomially with matrix size \cite{Shewchuk1994}, so the run-time increases exponentially with the dimension of the problem. In this paper, a quantum algorithm is used to produce a quantum state representing the normalized solution of the problem. Since this technique runs in polylog time, the curse of dimensionality can be broken. Thus, we can solve the linear system of equations based on the HHL algorithm \cite{HHL2009}. Our algorithm exploits properties of matrix $A$ to efficiently implement the HHL algorithm by simulating the unitary operator $e^{i A t}$. Though we are presenting an algorithm for the one-dimensional Poisson equation, this can be easily extended to the $d$-dimensional case \cite{Wang2020a, Liu2021} as
\begin{multline}
    A^{(d)} = \underbrace{A \otimes I \otimes \cdots \otimes I}_d 
    + I \otimes A \otimes I \otimes \cdots \otimes I
    + \cdots \\
    + I \otimes \cdots \otimes I \otimes A.
    \label{eq.4b}
\end{multline}

with the exponential $A^{(d)}$ expressed in the form
\begin{equation}
    e^{iA^{(d)}t} = \underbrace{e^{iAt} \otimes e^{iAt} \otimes \cdots \otimes e^{iAt}}_d.
    \label{eq.4c}
\end{equation}
So, the quantum circuit simulating $e^{iA^{(d)}t}$ is just the parallel execution of the circuit simulating $e^{iAt}$ along the $d$ dimension. In the following sections, we will focus on the one-dimensional Poisson equation. 

\begin{figure}[t]
\includegraphics[scale=1.18]{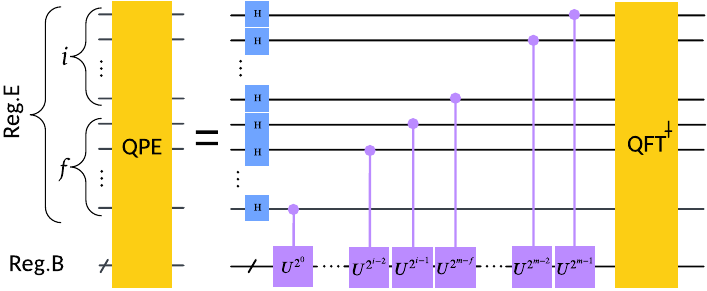}
\caption{Overall circuit for quantum phase estimation (QPE) that uses both integer and fractional parts of the eigenvalues through reg.\,E. $U^{2^k}$ represents the unitary operator of $\exp{(i2\pi A/2^{m-k})}$. QFT$^\dagger$ represents the inverse quantum Fourier transform.} 
\label{fig.2}
\end{figure}

\section{Quantum Algorithm and Circuit Design}
The overall circuit diagram of our algorithm for solving the one-dimensional Poisson equation is presented in Fig.\,\ref{fig.1} \cite{Wang2020a, Robson2022, Robson2022b}. As the figure shows, the algorithm consists of three stages: phase estimation, controlled rotation, and uncomputation. Its circuit diagram has three main registers -- reg.\,B, reg.\,E, and reg.\,A. 
\begin{itemize}
    \item Reg.\,B is used to encode the coefficients of the right-hand side of Eq.\,(1). Its number of qubits is $n=\lceil \log(N^\prime) \rceil$, where $N^\prime$ is defined in section \ref{overview}.
    \item Reg.\,E is used to store the approximated eigenvalues of matrix $A$. Its number of qubits is $m = i + f$, where the first $i = 2n + 2$ qubits hold the integer part and the remaining $f$ qubits the fractional part of the eigenvalue. 
    \item Reg.\,A is used to store pre-calculated angular coefficients for the controlled rotation operation. Its number of qubits is chosen to be $l \ge m$. 
\end{itemize}

In this work, we assume that the input state $|b\rangle$ of reg.\,B is prepared as $\sum_i b_i|i\rangle$, where $b_i$ is the value on the right-hand side of Eq.\,(3), and $|i\rangle$ is the computational basis \cite{Aaronson2015}. That is, the input $|b\rangle$ contains the prerequisite state vector, the problem that we are trying to solve, which we then entangle with the approximated eigenvalues \({\lambda}_j\) on reg.\,E. The output of the algorithm thus is a quantum state that encodes the solutions of the Poisson equation as probability amplitudes on reg.\,B. Thus, this circuit is a process of quantum state preparation, with the output written as $|v\rangle = A^{-1}|b\rangle = \sum_i\alpha_i|i\rangle$, where $\alpha_i$ is the value of the solutions of the Poisson equation after normalization.

We will now discuss a few key steps of the algorithm.
\subsection{Phase Estimation}\label{qpe}
Through the quantum phase estimation (QPE) circuit shown in Fig.\,\ref{fig.2}, we estimate the eigenvalues of the discretized matrix $A$ and entangle the states encoding the eigenvalues with the corresponding eigenstates \cite{Luis1996}. We will now discuss how the quantum states evolve through the QPE section of the circuit. The initial state of reg.\,E and reg.\,B is 
\begin{equation} \label{}
|0\rangle^{\otimes m}|b\rangle = \sum_{i=1}^{2^n - 1}b_i|0\rangle^{\otimes m}|i\rangle = \sum_{j=1}^{2^n - 1}\beta_j|0\rangle^{\otimes m}|u_j\rangle.
\label{eq.4}
\end{equation}
where \(|i\rangle\) is the computational basis and \(|u_j\rangle\) is the \(j\)th eigenvector of matrix \(A\). Then the Hadamard gates across reg.\,E prepare a uniform superposition state, which the sequence of controlled $U^{2^m}$ operation evolves as follows:
\begin{multline}
\sum_{k'=0}^{2^m-1}(|k'\rangle \langle k'| \otimes U^{k'}) \cdot {\frac {1} {\sqrt{2^m}}} \sum_{k=0}^{2^m-1}|k\rangle \otimes \sum_{j=1}^{2^n-1}\beta_j|u_j\rangle\\
= \sum_{j=1}^{2^n-1}\beta_j \left[ \frac{1}{\sqrt{2^m}} \sum_{k=0}^{2^m-1}e^{2\pi i \frac{{\lambda}_j}{2^m} k}|k\rangle \right]|u_j\rangle.
\label{eq.5}
\end{multline}

Note that the state in the square bracket of Eq.\,\ref{eq.5} is simply the output of the quantum Fourier transform acting on the state \(|{\lambda_j}\rangle\), so after the application of the inverse Fourier transform the states evolve to \(\sum_{j=1}^{2^n-1}\beta_j|{\lambda_j}\rangle|u_j\rangle\). This entangles the eigenvalues \(|\lambda_j\rangle\) with the eigenstates \(|u_j\rangle\) from reg.\,B.

Though there are methods \cite{Lloyd1996, Barry2007} available for simulating the time evolution of $e^{iAt}$, Wang et al. take advantage of using specific properties of the tri-diagonal matrix $A$ to reduce the complexity of the algorithm. They first decompose the unitary operator $e^{iAt}$ with a Hermitian matrix $S$ ($S$ being an orthogonal matrix composed of the eigenvectors of $A$) and then diagonalize it via the sine transform, and finally use phase kickback \cite{Cleve1998} to operate it on the state $|b\rangle$. We adopt Wang's approach for phase estimation; its detailed circuit composition is available in Ref. \cite{Wang2020a}.

\subsection{Phase Verification}
\begin{figure}
    \centering
    \includegraphics[scale=.25]{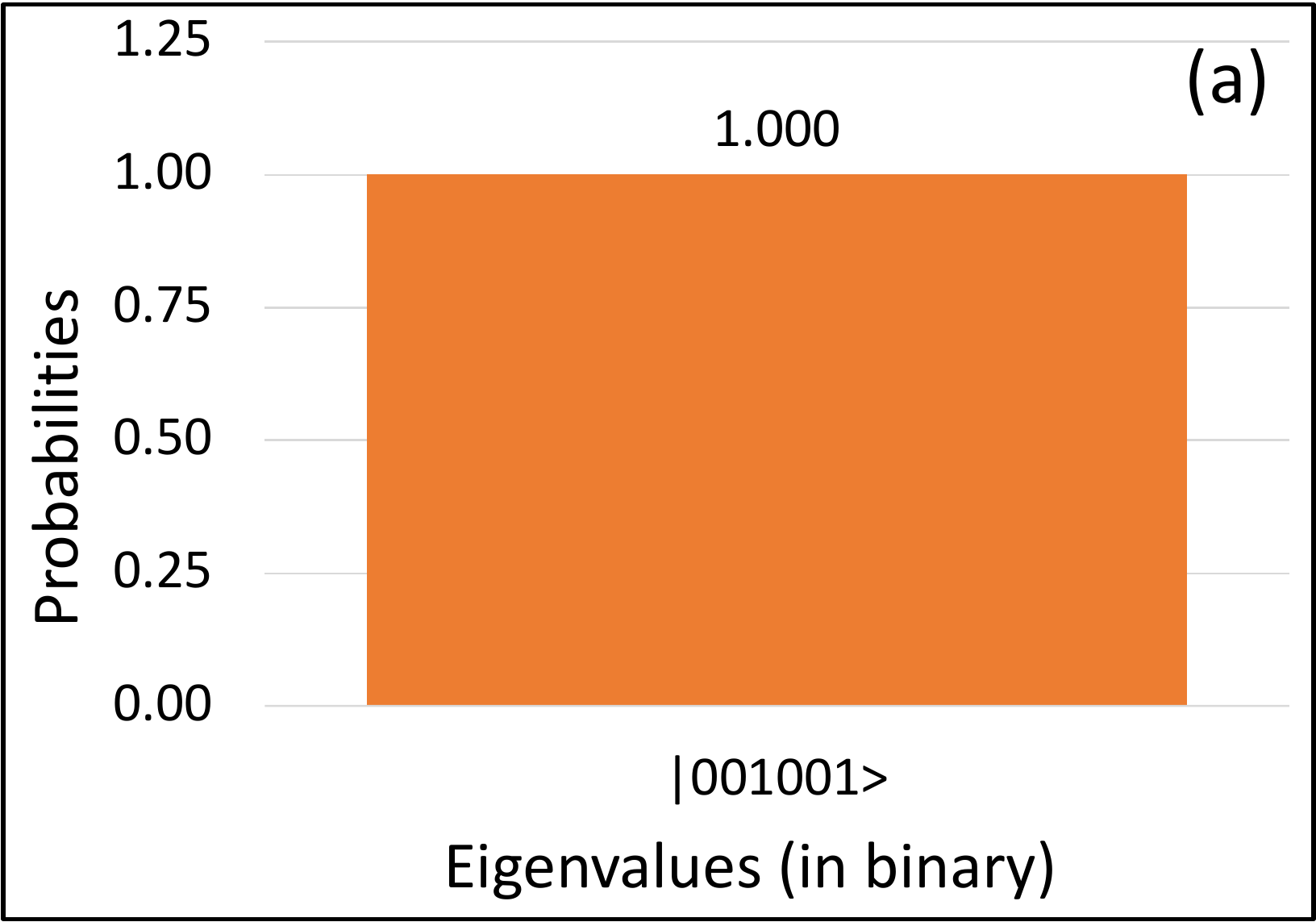}
    \includegraphics[scale=.25]{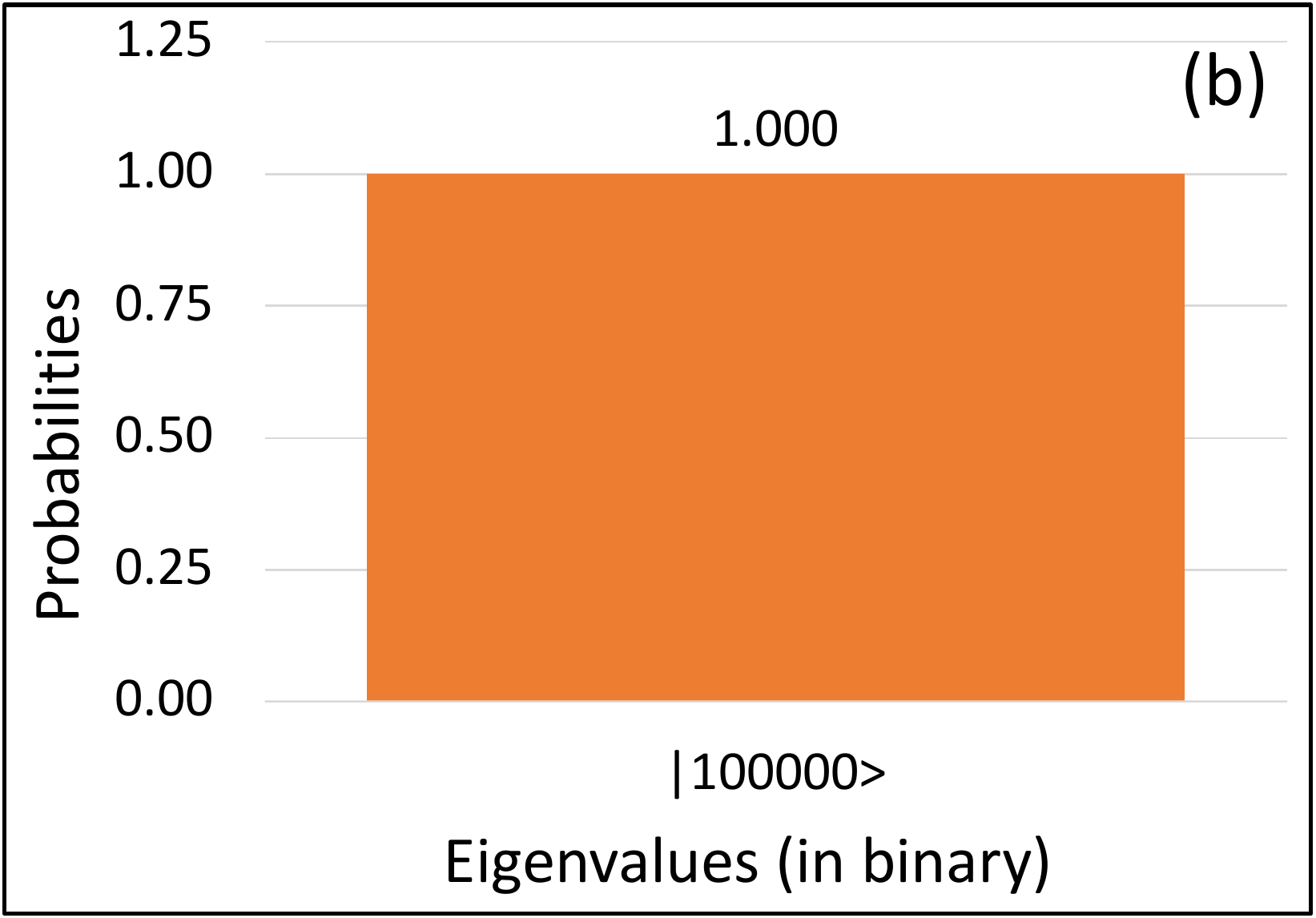}
    \includegraphics[scale=.25]{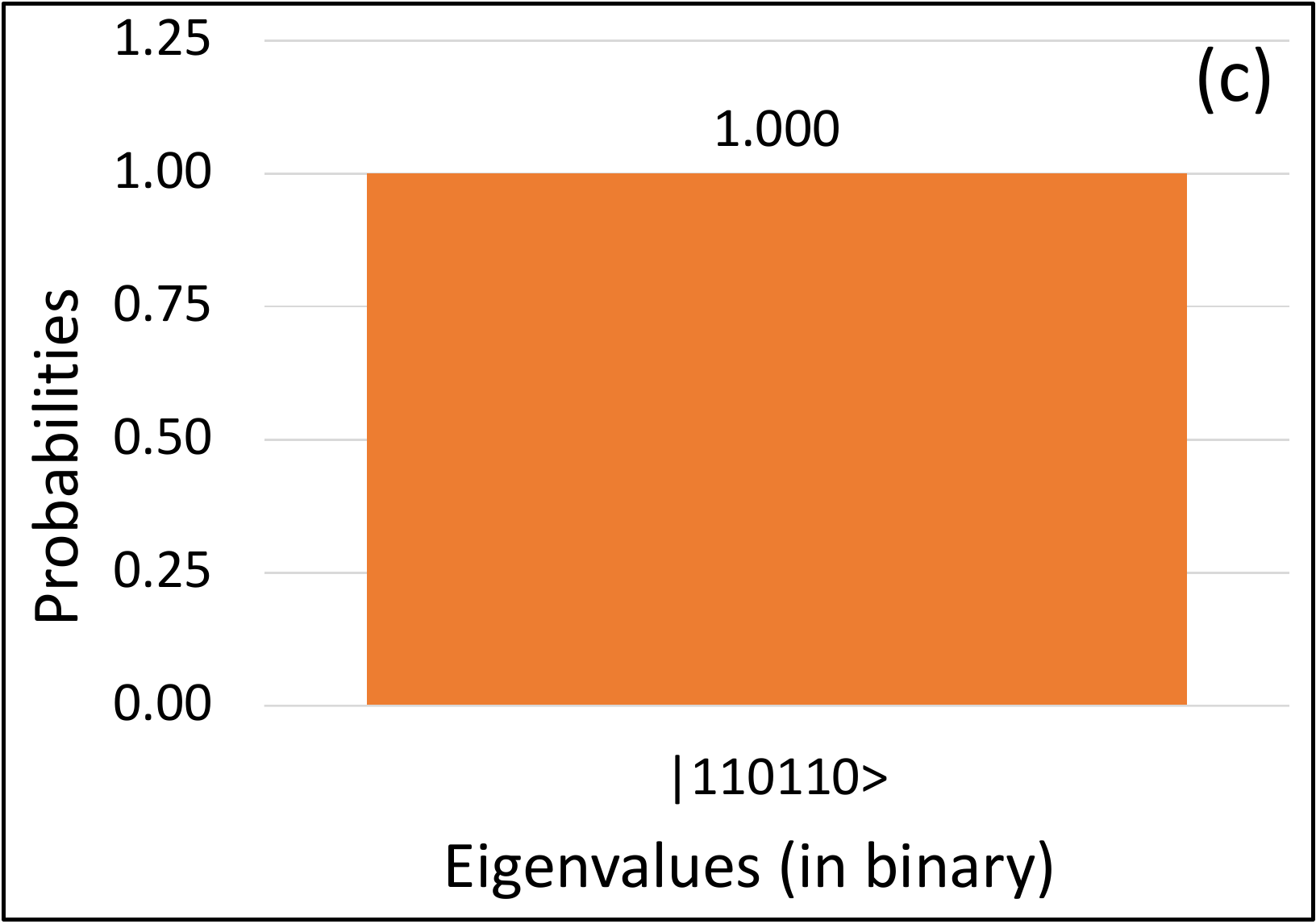}
    \includegraphics[scale=.25]{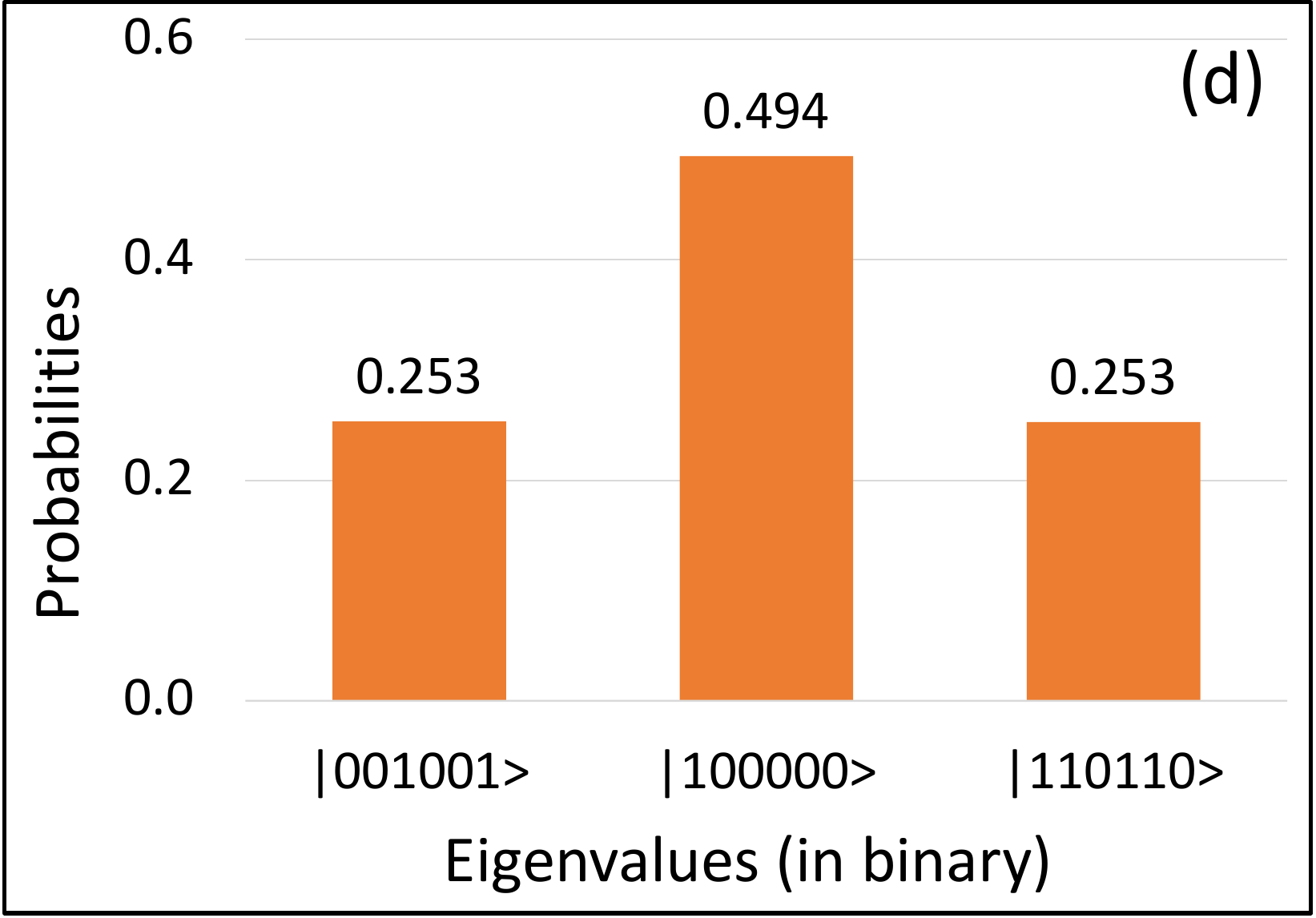}
    \caption{Verification of the QPE section of the circuit. Panels (a-c) show that for a given eigenstate of $A(3\times 3)$, the QPE produces the corresponding eigenvalue with $100\%$ probabilities. Panel (d) shows that inputting the combinations of eigenstates with arbitrary weights produces eigenvalues with similar weights.}
    \label{fig.3}
\end{figure}

An eigenvalue problem involving an arbitrary unitary operator $A$ and its eigenvector $|{v_j}\rangle$ and eigenvalue $\lambda_j$, satisfies $A|v_j\rangle=\lambda_j|v_j\rangle$. Using this, we can verify the correctness of the QPE part of the circuit. In fact, this would also implicitly verify the phase kickback operation, which, through the controlled $U$ operations (in Fig.\,\ref{fig.2}), entangles the eigenvalues of matrix $A$ with the eigenstates associated with the input in reg.\,B. We can think of reg.\,B as containing the problem we are trying to solve for the HHL algorithm. Each eigenvalue of matrix $A$ is associated with an eigenvector, so the first way to perform the verification is to input the individual eigenvectors as the input to reg.\,B, and then measure reg.\,E before the controlled rotations. For example, a Qiskit simulation with $A(3\times 3)$ in Eq.\,\ref{eq.3} acting on its eigenstates 
% $ |v_j\rangle =
% \begin{pmatrix}
% 1 \\
% \sqrt{2} \\
% 1 
% \end{pmatrix},
% \begin{pmatrix}
% -1 \\
% 0 \\
% 1 
% \end{pmatrix}, 
% \begin{pmatrix}
% 1 \\
% -\sqrt{2} \\
% 1
% \end{pmatrix}$ 
$ |v_j\rangle =
\left(\begin{smallmatrix}
1 \\
\sqrt{2} \\
1 
\end{smallmatrix}\right),
\left(\begin{smallmatrix}
-1 \\
0 \\
1 
\end{smallmatrix}\right), 
\left(\begin{smallmatrix}
1 \\
-\sqrt{2} \\
1
\end{smallmatrix}\right)$ 
produces the eigenvalues $\lambda_j = 9, 32, 54$ in binary (using only the integer part for simplicity) with $100\%$ probability; this is shown in Fig.\,\ref{fig.3}\,(a-c). Further, for any input with an arbitrary combination of eigenvectors, for example, $|v\rangle = \frac{1}{2}|v_1\rangle + \frac{1}{\sqrt{2}}|v_2\rangle + \frac{1}{2}|v_3\rangle$, the QPE circuit produces the combination of the eigenvalues with correct probabilities, as presented in Fig.\,\ref{fig.3}\,(d).

\subsection{Controlled Rotation}
After the phase estimation $\sum_{j=1}^{2^n-1}\beta_j|\lambda_j\rangle|u_j\rangle$ is obtained on regs.\,B and E, we perform the linear map taking the state of $|{\lambda}_j\rangle$ to $(1/\lambda_j) |\lambda_j\rangle$.
This process consists of two parts: calculating the rotation angular coefficients and performing the controlled $R_y$ operation. The probability amplitude of $1 / {\lambda}_j$ can be produced by implementing the
controlled $R_y$ rotation, that is, $R_y (2 \theta_j) |0\rangle = \cos{\theta_j} |0\rangle + \sin{\theta_j} |1\rangle$, 
where the rotation angle $\theta_j$ can be expressed in terms of of  ${\lambda}_j$ as 
\begin{equation} \label{}
		\sin{\theta_j} = 1/{\lambda}_j,
		\label{eq.6}
\end{equation}
which can be rewritten as
\begin{equation} \label{}
	\cot{\theta_j} = \sqrt{{\lambda}^{2}_j - 1}, ~~~ \theta_j \in (0, \pi/2).
	\label{eq.7}
\end{equation}
Taking $\theta_j = \omega_j \pi$, Eq.\,\ref{eq.7} becomes
\begin{equation} \label{}
	\omega_j = \frac{1}{\pi} {\rm arccot} \big(\sqrt{{\lambda}^{2}_j - 1} \big), ~~~ \omega_j \in (0, 1/2),
	\label{eq.8}
\end{equation}

where $\omega_j$ is the rotation angular coefficient. In this hybrid approach, we prepare $\omega_j$ classically and encode them into the circuit.

For an angular coefficient state $|\omega_j\rangle$ in reg.\,A, the binary representation can be written as $\omega_j = \omega_{j_1}\omega_{j_2}\cdots\omega_{j_l} = \sum_{k=1}^l 2^{-k}\omega_{j_k}$. Then using $R_y(2\theta_j) = e^{-i \theta_j Y}$, the $R_y$ rotation can be expressed as \cite{Wang2020c},
% \begin{equation}
\begin{multline}
    R_y(2\omega_j\pi) = e^{-i (\sum\limits_{k=1}^l\frac{\omega_{j_k}}{2^{-k}}) \pi Y} \\ 
    = \prod_{k=1}^l e^{-i \frac{\omega_{j_k}}{2^k}\pi Y} 
    = \prod_{k=1}^l R_y^{\omega_{j_k}} (\frac{\pi}{2^{k-1}}).
    \label{eq.9}
\end{multline}
% \end{equation}

where $\omega_{j_k}$ are the control qubits in reg.\,A. For a given $k$, if the bits of $\omega_{j_k}$ for all $j$ are zero, then the corresponding $R_y^{\omega_{j_k}}$ operation has no effect on the Ancillary register. This allows us to further optimize the circuit by removing any control qubits with bit $\omega_{j_k} = 0$ from reg.\,A. This is further discussed in the next section. 

% The algorithm used in this work follows several steps:
The workflow used in this work follows several steps and is presented in Algorithm~1.

\begin{algorithm}
\caption{Quantum Poisson Solver}\label{alg:cap}
% \SetKwInOut{Test1}{Test2} Test3
\KwData{Input state $\sum_{i}b_i |i\rangle$ in reg.\,B. For QPE, assign a number of qubits for the integer and fractional parts of $|\lambda_j\rangle$ in reg.\,E. Also, assign the initial number of qubits for reg.\,A for $|\omega_j\rangle$}
\KwResult{Get the solution $|v\rangle = A^{-1}|b\rangle$ in terms of probability amplitudes} 

Algorithm Start:
\begin{enumerate}
\item Prepare the initial quantum state: $\sum_{j=1}^{2^n - 1}\beta_j|0\rangle^{\otimes m}|u_j\rangle$

\item Use QPE algorithm on regs.\,B and E. This algorithm applies several Hamiltonian simulations of $U = e^{i A t}$ with $t = 2\pi \frac{1}{2^n} 2^k$, \, $k=0,..., n-1$, to reg.\,B and entangles the eigenvalues $\lambda_j$ of matrix $A$ in reg.\,E with the eigenstates $|u_j\rangle$ in reg.\,B. The system has now the state: $\sum_{j=1}^{2^n - 1} \beta_j  |{\lambda}_j\rangle |u_j\rangle$

\item Apply the controlled rotation which consists of two parts: preparing the rotation angular coefficients $|\omega_j\rangle$ in reg.\,A and performing the controlled $R_y(2\omega_j \pi)$ operation on the ancillary qubit
	
\item Uncompute QPE and $|\omega_j\rangle$ operations on regs.\,A, E and B
	
\item Measure the ancillary qubit. If the measurement of the qubit results in state $|1\rangle$, the algorithm successfully transforms reg.\,B into the solution $|v\rangle = A^{-1}|b\rangle = \sum_{j=1}^{2^n - 1} \beta_j \frac{1}{{\lambda}_j} |u_j\rangle$. Otherwise, the algorithm has to be restarted

\item Take a repeated number of trials for a good sampling  

\item Take the sum of individual successful states $|v_i\rangle$ and derive final probability amplitudes as $\sqrt{v_i/\sum_i v_i}$  
\end{enumerate}
\end{algorithm}

\section{Algorithm Improvements, Circuit Optimization and Challenges}

Wang's method has already reduced its complexity to $O(m^2)$ qubits and $O(m^3)$ operations \cite{Wang2020a}. After implementing the algorithm as a quantum circuit, we look for options for further improving it so that even with a limited number of qubits on the quantum hardware, we are able to more accurately solve the Poisson equation for a realistic problem size, i.e., with a larger matrix $A$. Below, we discuss some shortcomings of the existing approaches and the ways we improve them: 

\subsection{Eigenvalue Amplification}\label{amplification}
The first source of inaccuracy in the existing algorithm \cite{Cao2013, Wang2020a} is the truncation of the eigenvalues of matrix $A$ in the phase estimation. Wang's implementation uses only the integer eigenvalues of the $A$ matrix, presumably in order to save qubits. As more qubits become available in quantum devices, however, we can improve accuracy by using non-truncated values. Therefore, we extend the algorithm by taking into account both the integer and fractional parts of the eigenvalues. This is done by amplifying the eigenvalue by a factor of $2^f$, which shifts the decimal point of the binary $\lambda_j$ to right by an integer $f$. For example, for a given $\lambda_j = 10111.11011011101011$, with no amplification, $2^4$ amplification and $2^8$ amplification, the circuit carries $\lambda_j = 10111, 101111101$, and $1011111011011$, respectively. Essentially, when we include fractional part for the eigenvalue, we are encoding a bitshifted/amplified eigenvalue that is still an integer but contains bits of the fractional part. This way, by using a large $f$, one actually includes more bits of the fractional part of the eigenvalue, and the shifted position of the decimal point of the eigenvalue is adjusted by a normalization factor $2^{-f}$ in the controlled $R_y$ operation of the circuit to match. Due to the dynamic nature of our code, we are able to experiment using any number of bits on the eigenvalues, taking a more accurate representation of the critical matrix $A$ for our computation.  

\subsection{Rotation Angular Coefficient Accuracy}
The second source of inaccuracy is in the calculation of the rotation angular coefficient. The previous method \cite{Wang2020a} omitted the subtrahend 1 under the square root in Eq.\,\ref{eq.8}; we instead include it. Additionally, we retain full accuracy in the calculation of the rotation angular coefficients by using the full eigenvalues. Furthermore, our implementation allows us to dynamically expand the number of qubits to represent rotation angular coefficients with higher accuracy. Finally, we are able to use the optimum number of qubits based on the convergence of the error in the solution, which is discussed in the next section. 

\subsection{Optimize the Number of Qubits Used for Rotation Angles}
While we initially presented a rotation on all bits of reg.\,A, in practice, this is not necessary. In Eq.\,\ref{eq.9}, if the bits of $\omega_{j_k}$ for all $j$ are zero for a given $k$, then the corresponding $R_y^{\omega_{j_k}}$ operation has no effect on the Ancillary register. In other words, if there is no information conveyed on a given qubit in reg.\,A by any of the rotation angular coefficients, then we can safely omit that qubit and its rotation without impacting the results. Intuitively, this makes sense as the controlled rotations do not happen if a given control bit is 0. Imagine a case where our $\omega_j = 0.0000100110, 0.0000001010, 0.0000000101$. The first four qubits, as well as the sixth qubit, are 0 for all $\omega_j$, so the respective $R_y(2^{-1}\pi)$, $R_y(2^{-2}\pi)$, $R_y(2^{-3}\pi)$, $R_y(2^{-4}\pi)$, and $R_y(2^{-6}\pi)$ rotations never happen. We have no need to include these controlled rotations nor the qubits in reg.\,A that they correspond to. This allows us to further optimize the circuit by removing any control qubits with bit $\omega_{j_k} = 0$ from reg.\,A. As a result, though at the beginning we chose $l\ge m$ qubits for reg.\,A, after the circuit optimization, the register has fewer than $l$ qubits. 

\subsection{Optimize \textsl{CNOT} Gates Usage}
The rotation angular coefficient allows us to entangle the prepared state on reg.\,E with the controlled rotation on reg.\,A. This is achieved with multi-controlled multi-target (MCMT) gates controlled on the binary expansion of the eigenvalues on reg.\,E. However, MCMT gates transpile to many \textsl{CNOT} gates, which carry significant errors into the experiment. We want to minimize the number of controlled bits in this operation. At the end of the phase estimation, the qubits of reg.\,E are entangled, thus the phase information can be accessed through fewer qubits in reg.\,E. This allows us to control our encoding of the rotation angular coefficients on only the unique most significant bits of reg.\,E. For example, in the $3 \times 3$ case, if our eigenvalues are 9, 32, 54 (taking only the integer part for simplicity), then their binary encodings are 001001, 10000, 110110, respectively. It is then evident that the two most significant bits of the binary encodings are enough to differentiate between the different eigenvalues: 00, 10, and 11. Controlling the angular rotations on only these two qubits allows us to reduce the number of \textsl{CNOT} gates in the circuit significantly. 

\subsection{Classical vs. Quantum Approach to Rotation Angular Coefficient}

Preparing the rotation angular coefficient $\omega_j$ using a quantum circuit has a cost that grows exponentially with the problem size \cite{Wang2020a}. This proves to be a challenge because the current state of simulator and quantum hardware supports a limited number of qubits. Therefore, though from a theoretical standpoint the quantum approach to $\omega_j$ is appealing, from a practical standpoint its classical treatment is the viable option. This is particularly true because our main goal is to scale the Poisson solver to realistic problem sizes, which requires us to appropriately allocate computational resources.

Therefore, in this work, we pre-calculate $\omega_j$ classically and encode them into the circuit. Though these two steps of the workflow are already fast, one can make them much faster by parallelizing them over $j$ on CPU or GPU hardware. Please note that this computation is required only once, independent of the number of repeated shots, and we make the process substantially more efficient by dynamically calculating it for any problem size.  

However, even within this hybrid approach, dealing with very large problems, e.g., encoding $10^8$ values of $\omega_j$ for a $10^8\times 10^8$ matrix, would be challenging. Such a large problem would make the circuit depth prohibitively large from an experimental standpoint. A multi-level solution to this problem is discussed in the subsequent section.

% \section{Circuits Demonstrated on Quantum Simulator}
\section{Simulated Results and Discussions}
We constructed our algorithm in the Python programming language using IBM's Qiskit package \cite{Qiskit}. This allowed us to create our circuit in a modular fashion as well as use some of Qiskit's abstractions, such as the MCMT gate and simple implementations of quantum Fourier transform.

\begin{figure}[h]
    \centering
    \includegraphics[scale=.275]{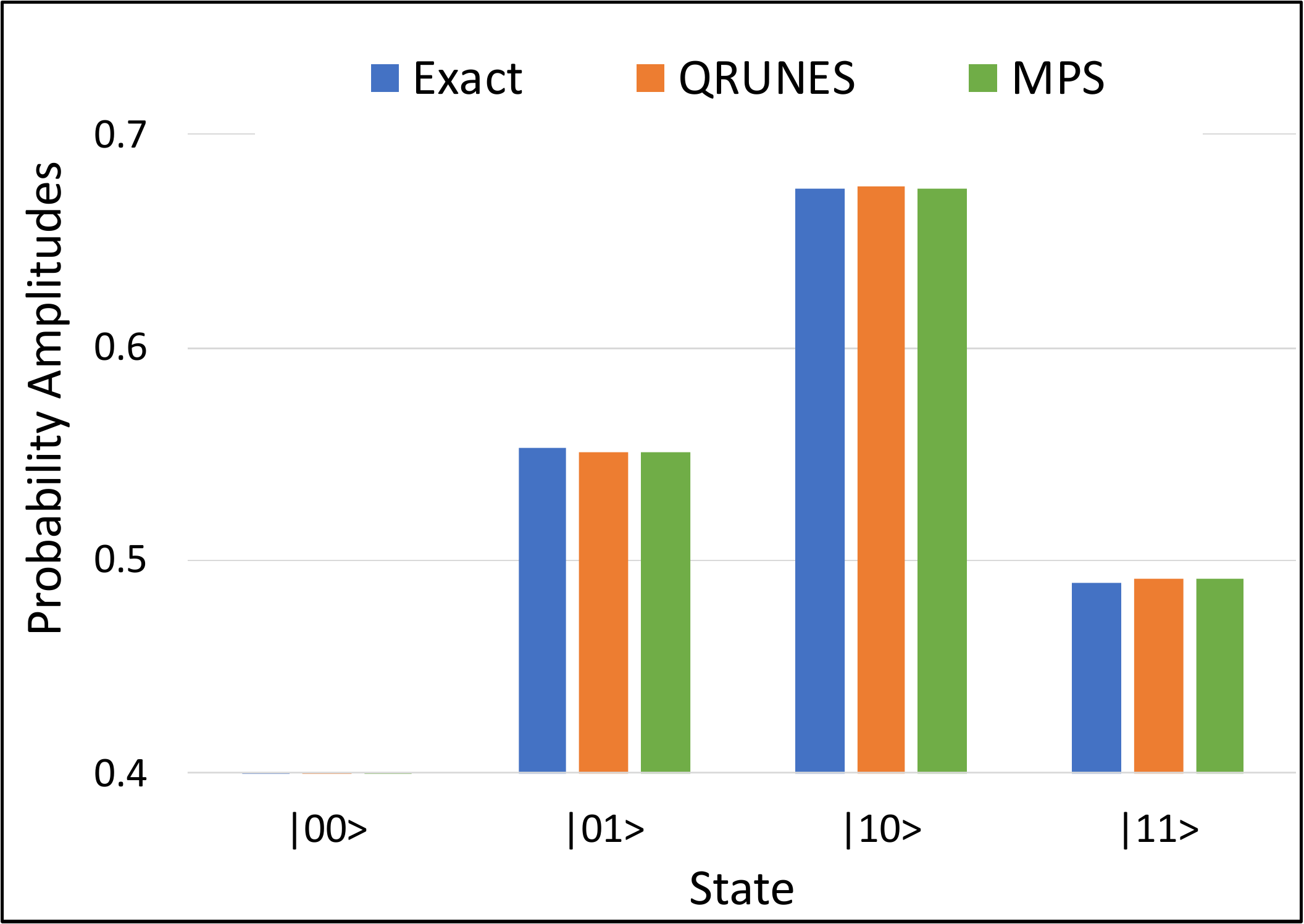}
    \includegraphics[scale=.275]{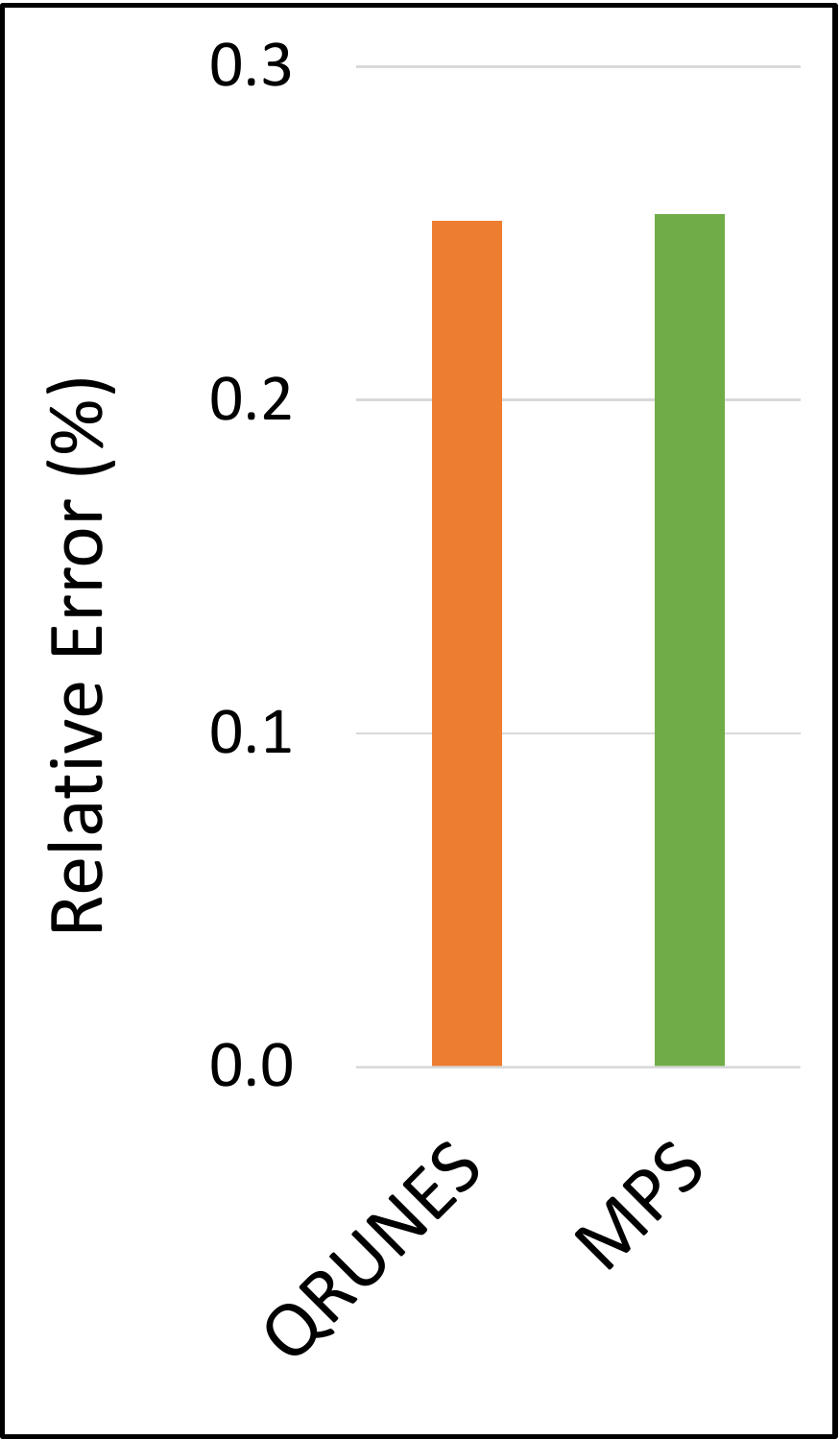}
    \caption{(Left) Comparison of the Matrix Product State (MPS)-based simulated solution of the Poisson equation with exact and existing QRUNES \cite{Wang2020a} solutions for a $3 \times 3$ problem size. (Right) The relative error in QRUNES and MPS simulated outputs with respect to the exact result.}
    \label{fig.4}
\end{figure}

% \begin{multicols}
\begin{figure*}
    \centering
    \includegraphics[scale=.26]{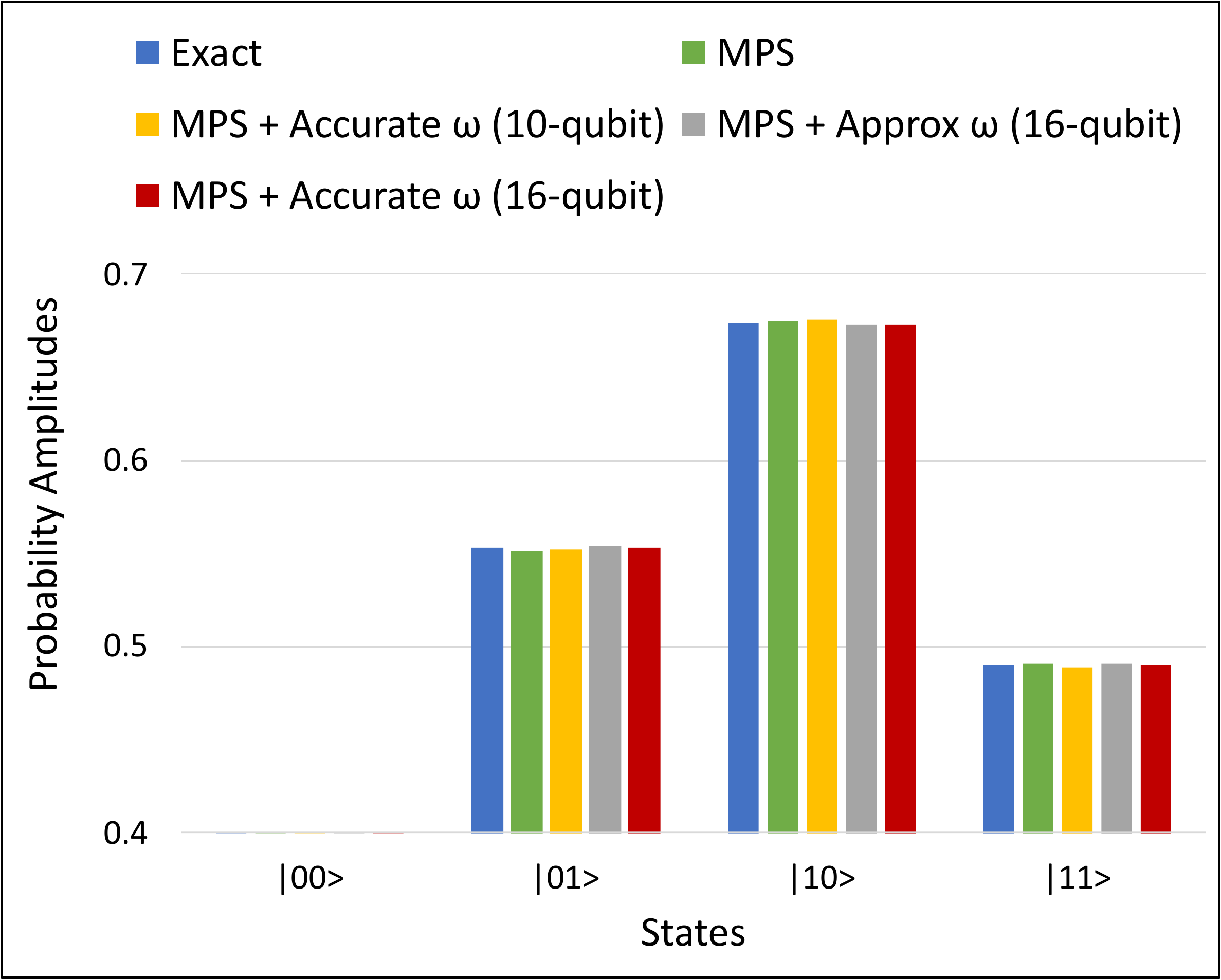}
    \includegraphics[scale=.26]{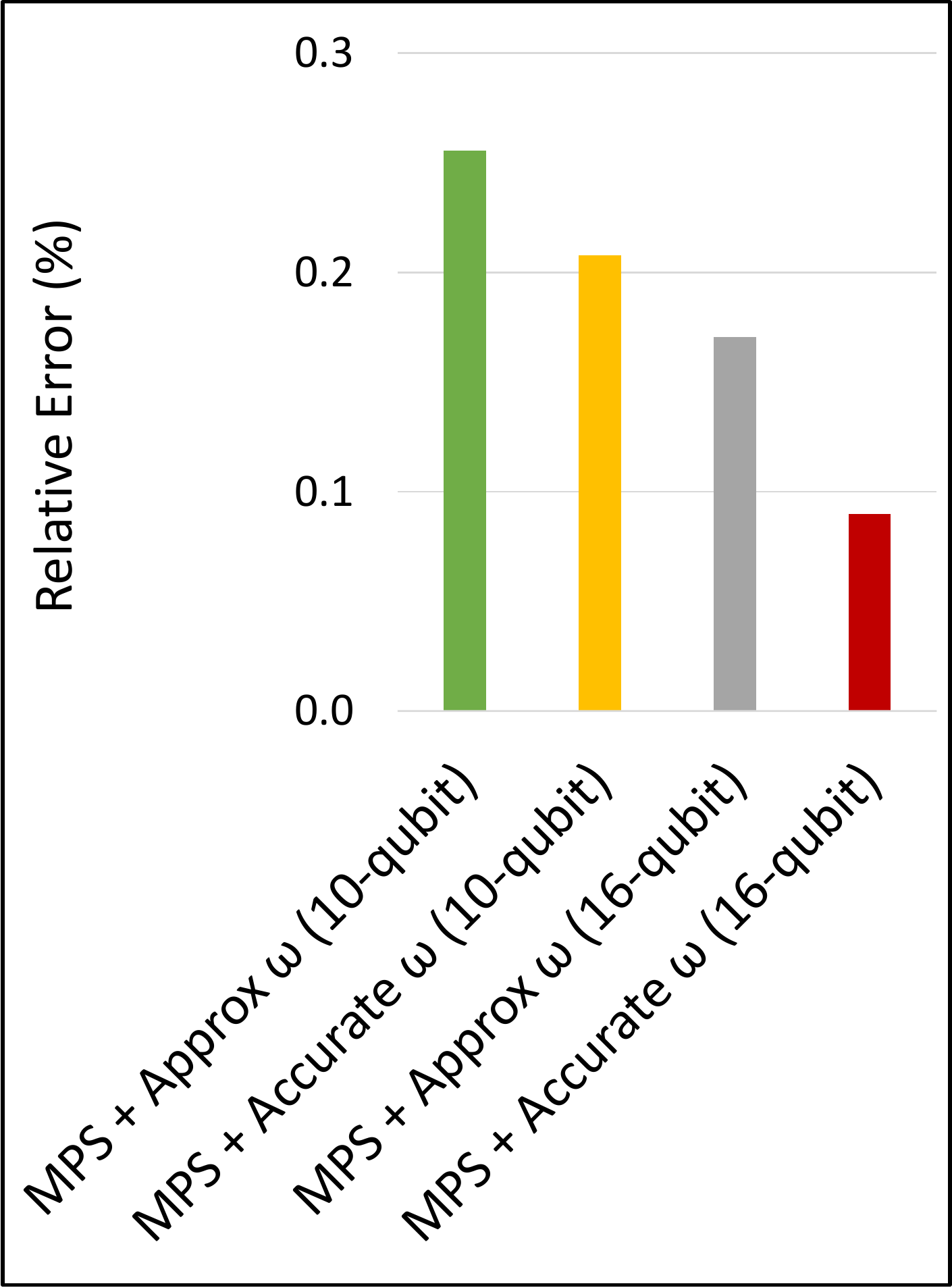}
    \includegraphics[scale=.26]{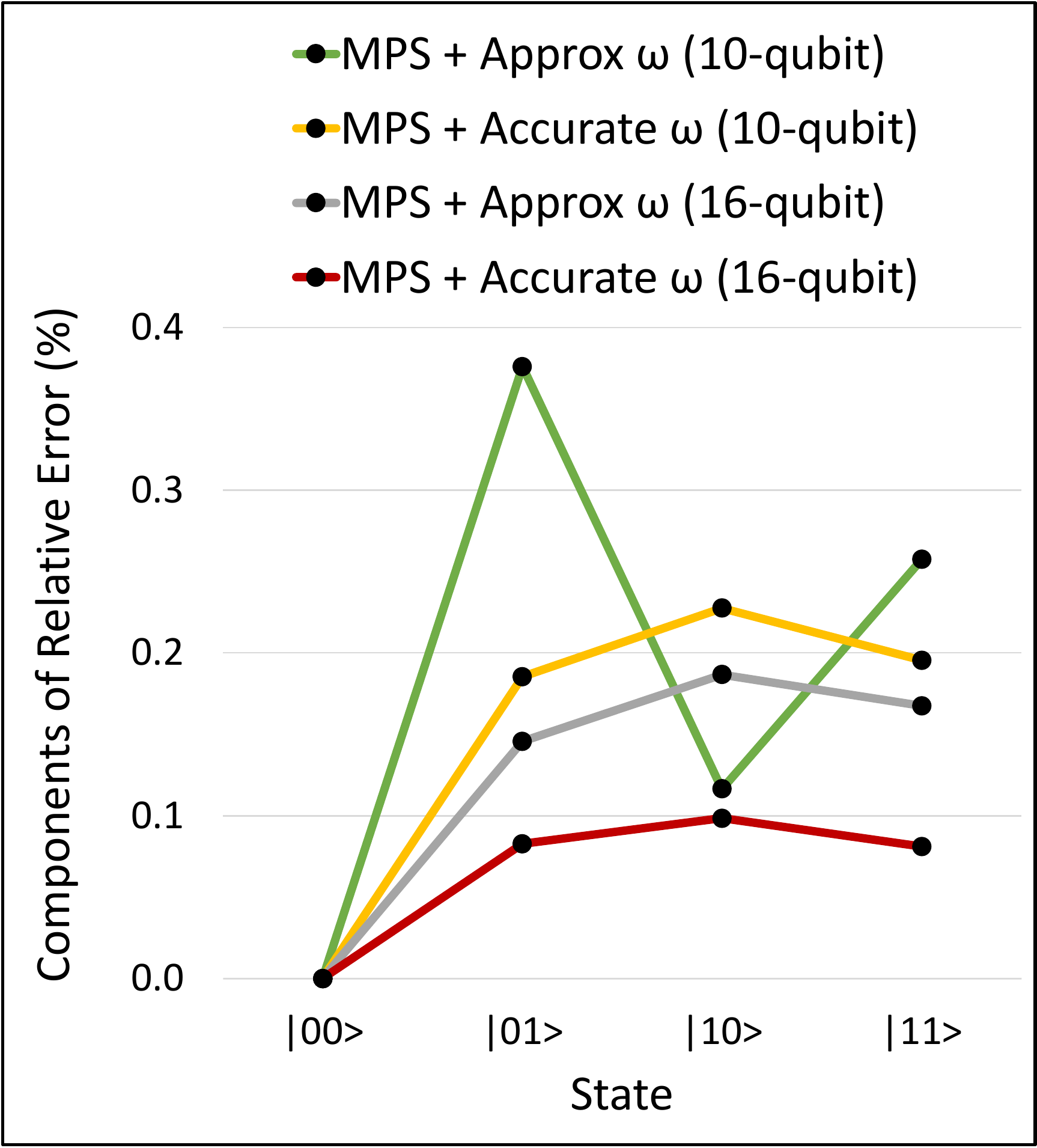}
    \caption{(Left) Shows step-by-step improvement in MPS-simulated solution by using accurate rotation angular coefficients $\omega_j$ and encoding them into up to 16 qubits on reg.\,A. (Middle) Shows the relative error in the improved MPS-simulated solutions with respect to the exact result. (Right) Shows relative errors at the level of individual states.}
    \label{fig.5}
\end{figure*}
% \end{multicols}

To the best of our knowledge, previous work has not included the simulation of a solution of the one-dimensional quantum Poisson equation beyond a $3\times 3$ matrix $A$.  However, here we present the simulation of solutions of much larger problems, that is, for larger sizes of $A$. In fact, we will present that our algorithm and its circuit representation are capable of dynamically controlling problem size in NISQ devices. For simulation, we use IBM's Matrix Product State (MPS) simulator since it supports a relatively large number of qubits (up to 100) necessary for presenting the circuit for larger problems while also maintaining reasonable accuracy. In this section, we analyze the source of error in the solution and accordingly demonstrate step-by-step improvements in the algorithm that secure higher accuracy in the solution.

\textsl{Reproduce Existing Results.} As shown in Fig.\,\ref{fig.4}\,(left), we first produce the solution of a $3 \times 3$ problem with $|b\rangle = \frac{1}{\sqrt{2}}|01\rangle + \frac{1}{2}(|10\rangle + |11\rangle)$ being the right-hand side of the Poisson equation. In order to compare this with the existing QRUNES results \cite{Wang2020a}, we use their same inputs, that is, only the integer part of $\lambda_j$ and the approximated $\omega_j$ encoded on 10 qubits of reg.\,A. Notice that in Fig.\,\ref{fig.4}\,(left), we show the vertical axis starting from 0.4, so that even any tiny differences in the heights of the histograms are clearly visible. Though our MPS-based simulated solution shows an excellent agreement with QRUNES, there are some discrepancies compared to the exact solution. To analyze further, the accuracy of our MPS-based result is depicted using the relative error in the MPS solution with respect to the exact result and, here the relative error, e.\,g.\,, for MPS is defined as ${\lVert\text{Exact} - \text{MPS}\rVert}_2/{\lVert\text{Exact}\rVert}_2$ \cite{Errors} and shown in the right panel of Fig.\,\ref{fig.4}. Relative errors in both QRUNES and MPS are virtually equivalent.

\textsl{Improvements in Results.} To improve our MPS-based result presented above and have a better agreement with the exact solution, we made the following two improvements: First, we used the accurate formula for $\omega_j$ given in Eq.\,\ref{eq.8}, then we encoded these values in up to 16 qubits on reg.\,A. The results are shown in the left panel of Fig.\,\ref{fig.5}, which displays the gradual improvements in the solutions as compared to the exact result. The improvements in solutions are clearly visible through the relative error presented in the middle panel of Fig.\,\ref{fig.5}, and its right panel explicitly shows the components of those relative errors.

\begin{figure}[h]
    \centering
    \includegraphics[scale=.275]{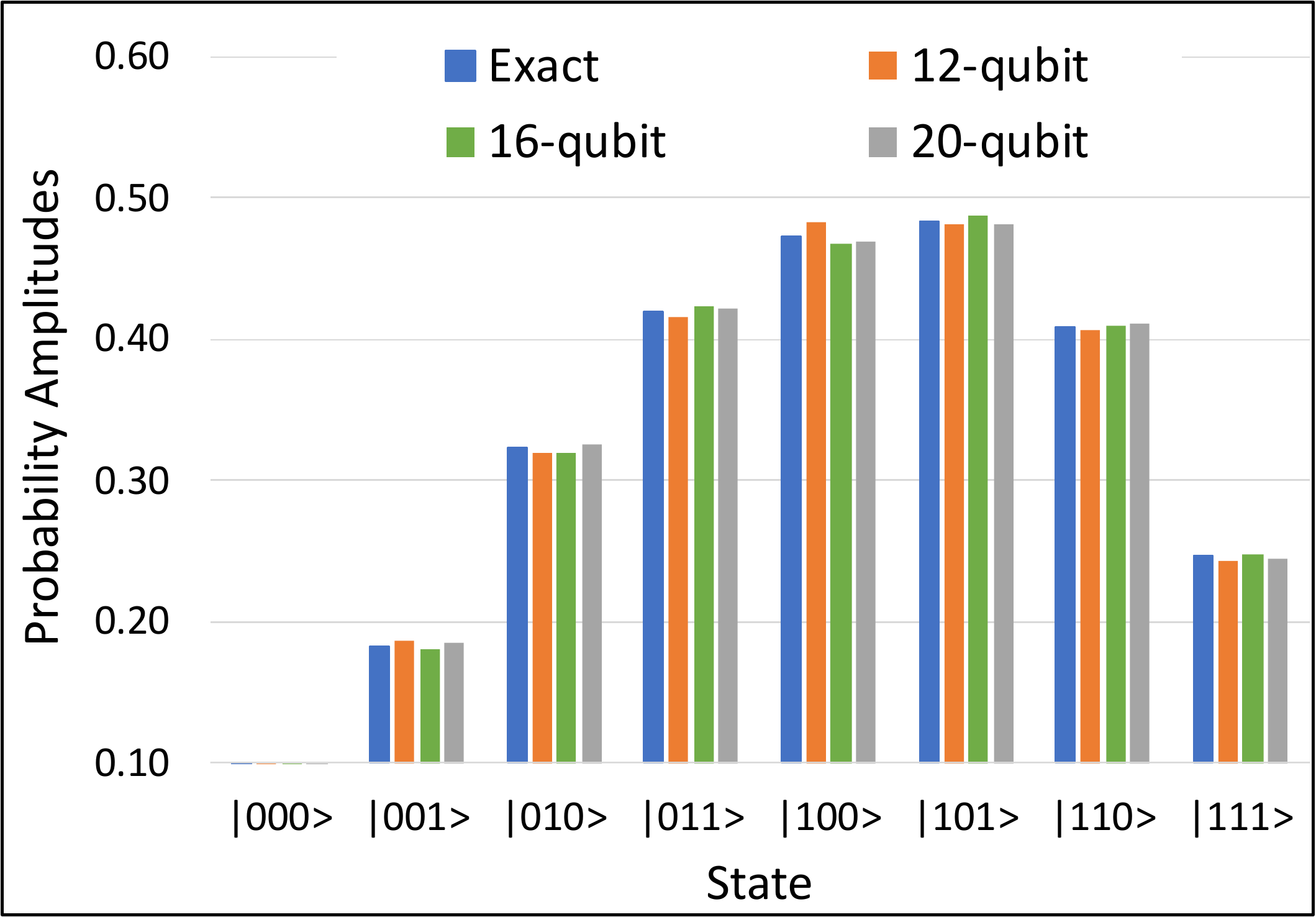}
    \includegraphics[scale=.275]{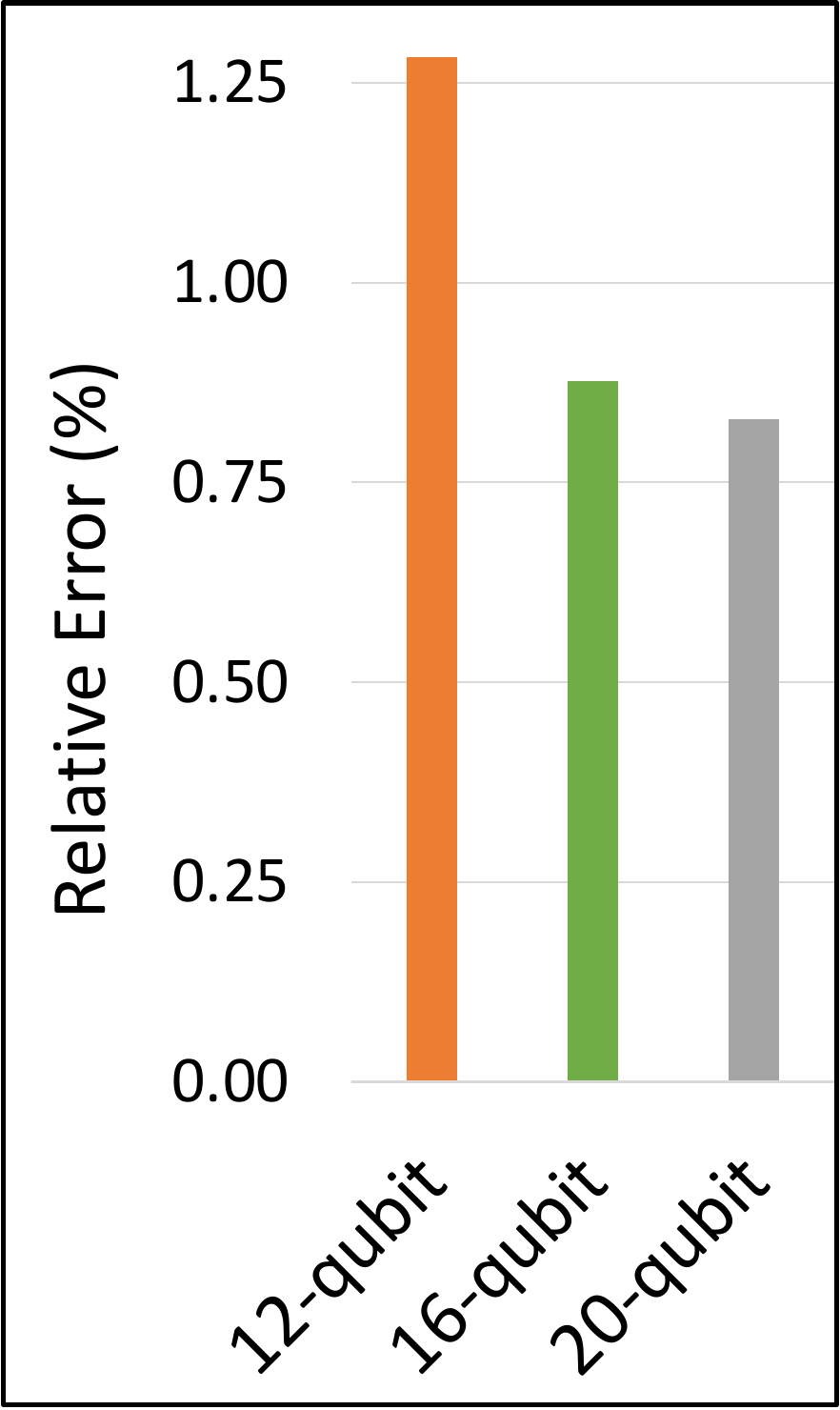}
    \caption{Effects of increasing the number of qubits on the rotation angular coefficients $\omega_j$ on a $7 \times 7$ problem size. (Left) Compares the exact solution with the MPS result simulated by encoding $\omega_j$ on different numbers of qubits of reg.\,A. (Right) Relative errors of the left panel results with respect to the exact solution.} 
    \label{fig.6}
\end{figure}

Next, we extend the problem size to $7 \times 7$ with $|b\rangle = \frac{1}{4}(|001\rangle + |010\rangle + |011\rangle + |100\rangle) + \frac{1}{2}(|101\rangle + |110\rangle + |111\rangle)$ and further investigate the effects of using a more precise $\omega_j$ by increasing the number of qubits in reg.\,A. As clear in Fig.\,\ref{fig.6}, improvements in the solutions and reduction of the relative errors resulted as the qubit number increased from 12 to 20. The relative error of the $7 \times 7$ problem with $\omega_j$ encoded in 16 qubits is about $0.88\%$, which is about $9$ times larger than that of the $3 \times 3$ problem. This is understood by the fact that the error due to the truncated eigenvalues $\lambda_j$ used in phase estimation plays a major role here. This is because a larger problem requires a larger number of controlled-$U$ operations (see Fig.\,\ref{fig.2}), resulting in more error accumulation. The overall accuracy of the results is relatively similar when using 16 and 20 qubits, thus we chose to fix $\omega_j$ at 16 qubits as we investigated further improvements to the solutions.

\begin{figure}[t]
    \centering
    \includegraphics[scale=.275]{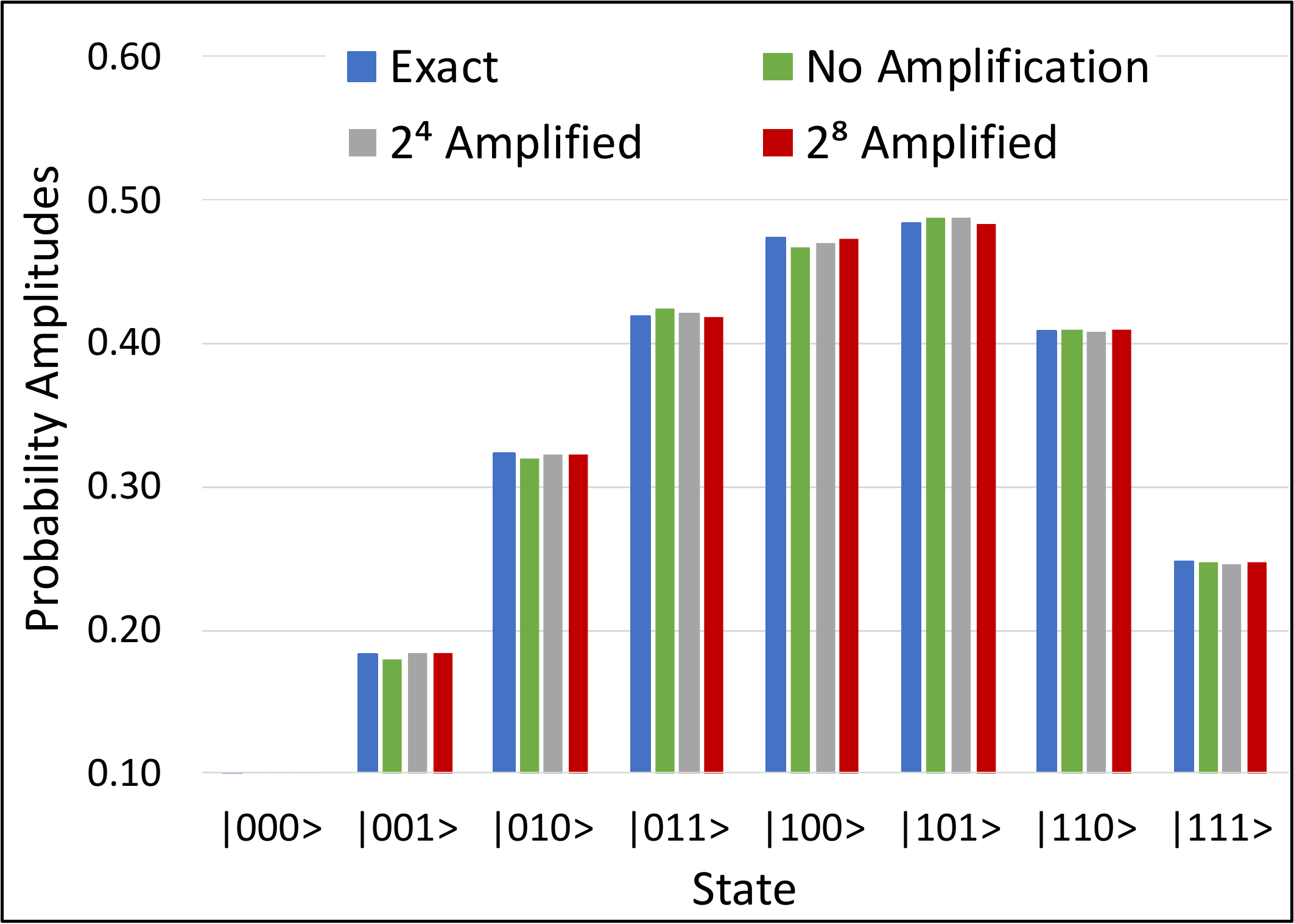}
    \includegraphics[scale=.275]{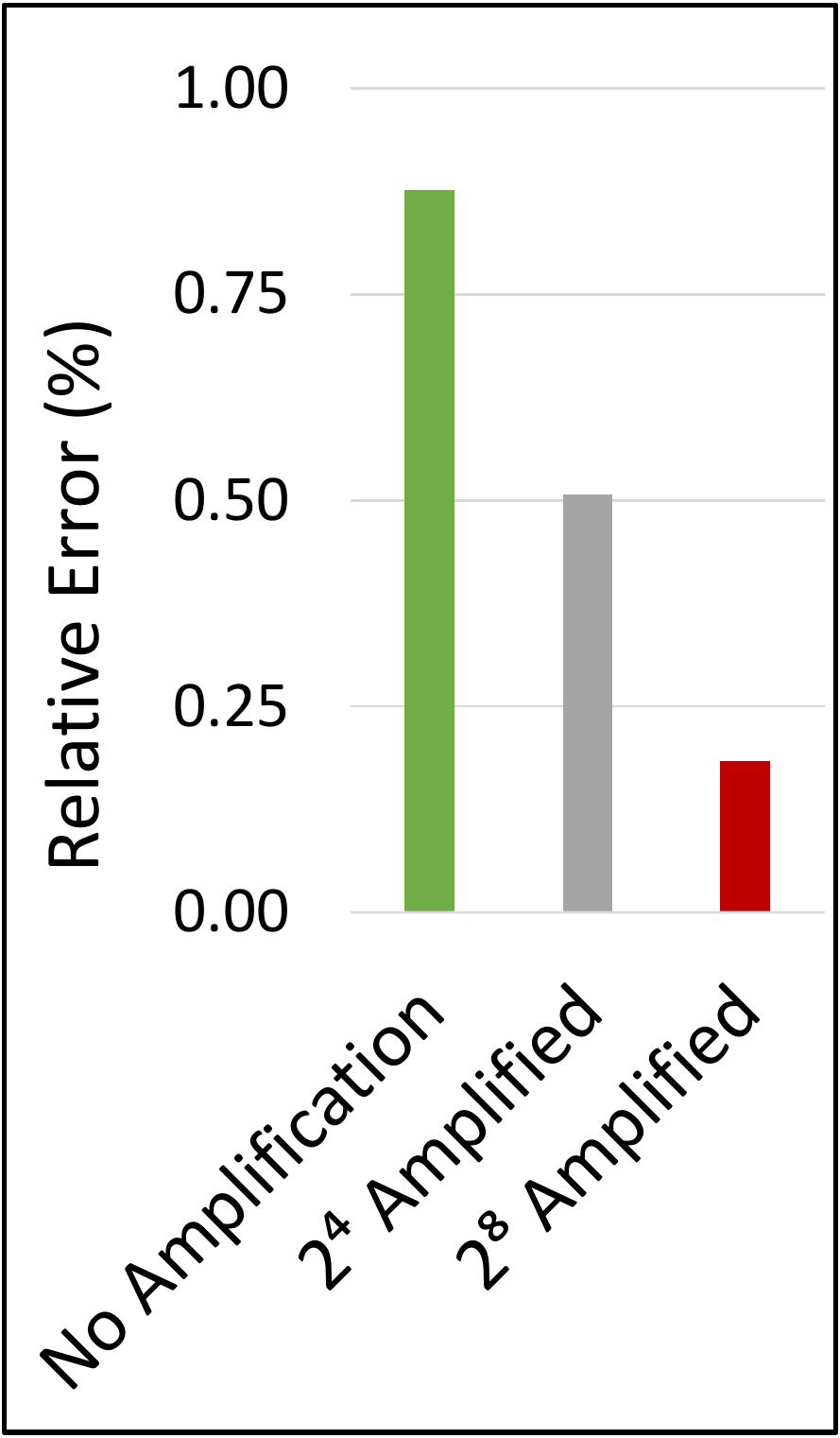}
    \caption{Effects of eigenvalue amplification on a $7 \times 7$ problem size. (Left) Comparing solutions with varying levels of eigenvalue amplification while using 16 qubits for $\omega_j$. (Right) Shows the drastic reduction in relative errors of the solutions on the left panel with respect to the exact result.}
    \label{fig.7}
\end{figure}

To further reduce the error discussed in the previous paragraph, we used eigenvalue amplification (as discussed in section \ref{amplification}) with a factor of $2^f$ where $f$ takes the value $0, 4$, and $8$. A larger $f$ includes more number of bits in the fractional part of the eigenvalue, and thus retains more accuracy in the solution. The effects of eigenvalue amplification on the $7 \times 7$ problem is shown in Fig.\,\ref{fig.7}, which confirms the significant reduction to the relative error when we use eigenvalue amplification. At $2^8$ amplification, the relative error is $0.18\%$, a 5-fold improvement in accuracy compared to using no amplification. We are confident that a higher amplification factor ($f>8$) would further reduce this error.

\begin{figure*}
  \includegraphics[scale=.35]{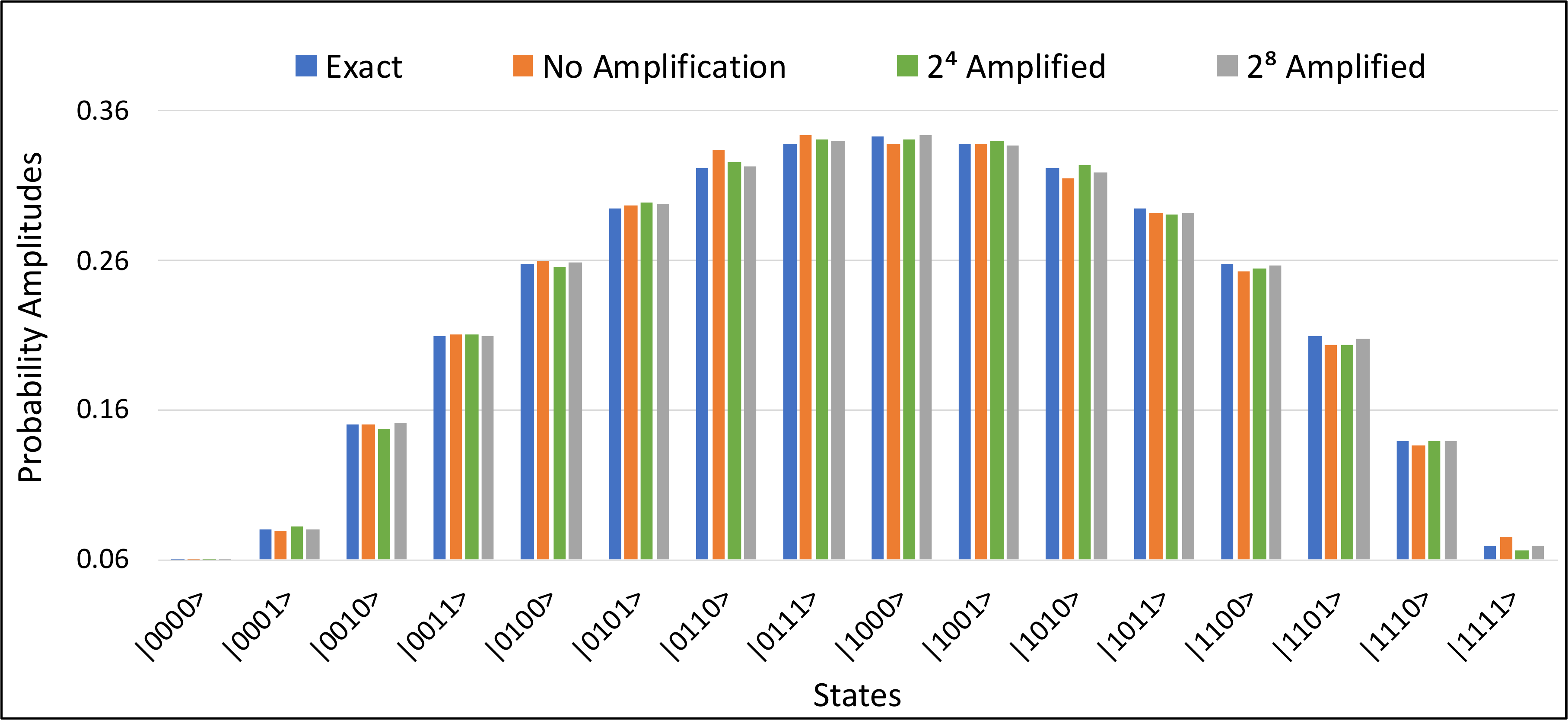}
  \includegraphics[scale=.35]{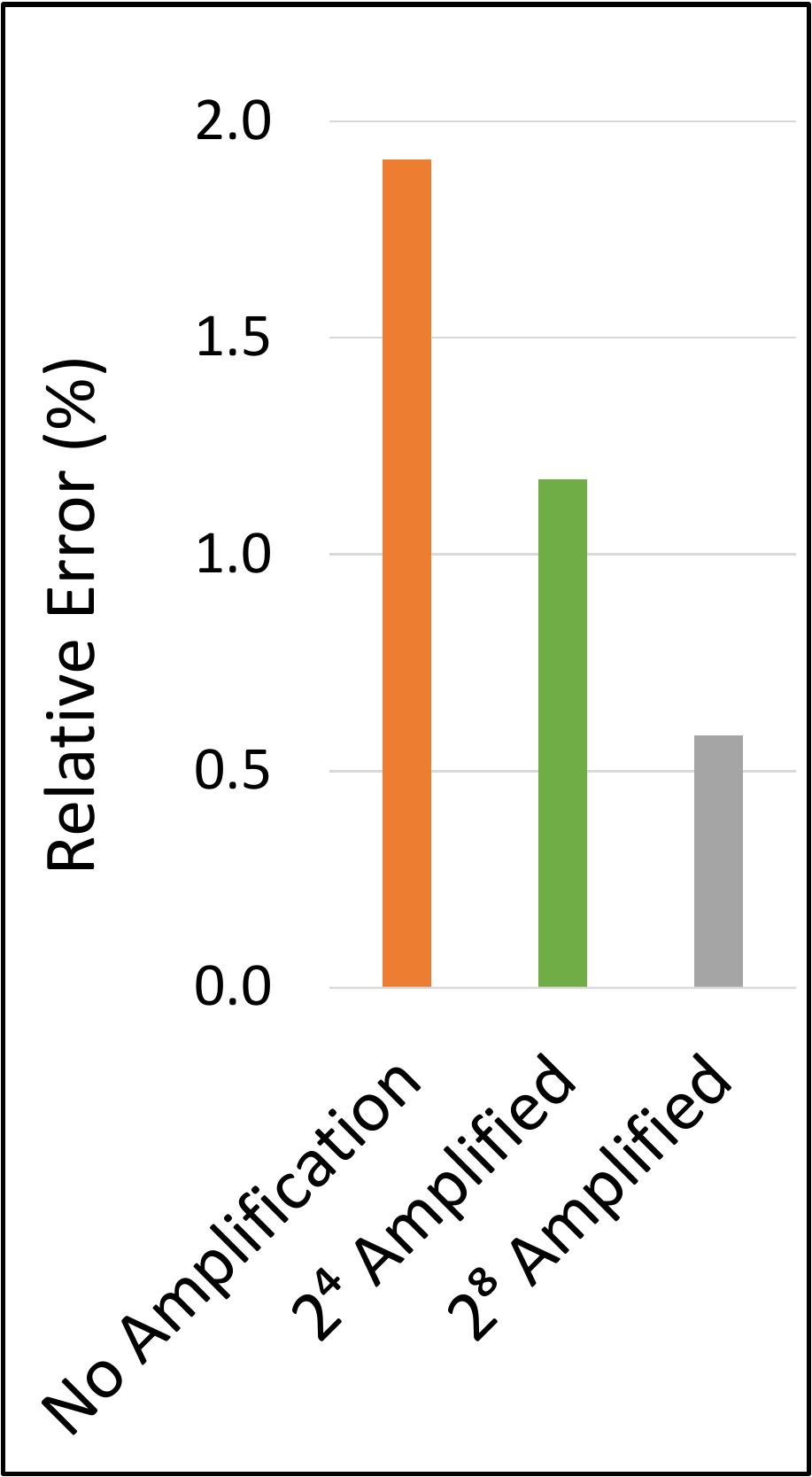}
  \caption{Shows solution of a $15 \times 15$ problem size. (Left) Comparing the solutions with different levels of eigenvalue amplification (with a fixed $\omega_j$ of 16 qubits) and exact result. (Right) Shows a significant reduction of the relative errors with respect to the exact result.}
  \label{fig.8}
\end{figure*}

To confirm the robustness of our algorithm and its accuracy in solving the Poisson equation for practical problem sizes, we present the solution for a $15 \times 15$ problem, including its exact result, in Fig.\,\ref{fig.8}. For an arbitrarily chosen input state $|b\rangle$ (see Table 1 for its expression), the overall solution is encouraging. The relative error with respect to the exact result again quickly goes down as we apply eigenvalue amplification and increase its amplification factor.

% \begin{multicols}
% \end{multicols}

\begin{figure}
    \centering
    \includegraphics[scale=.272]{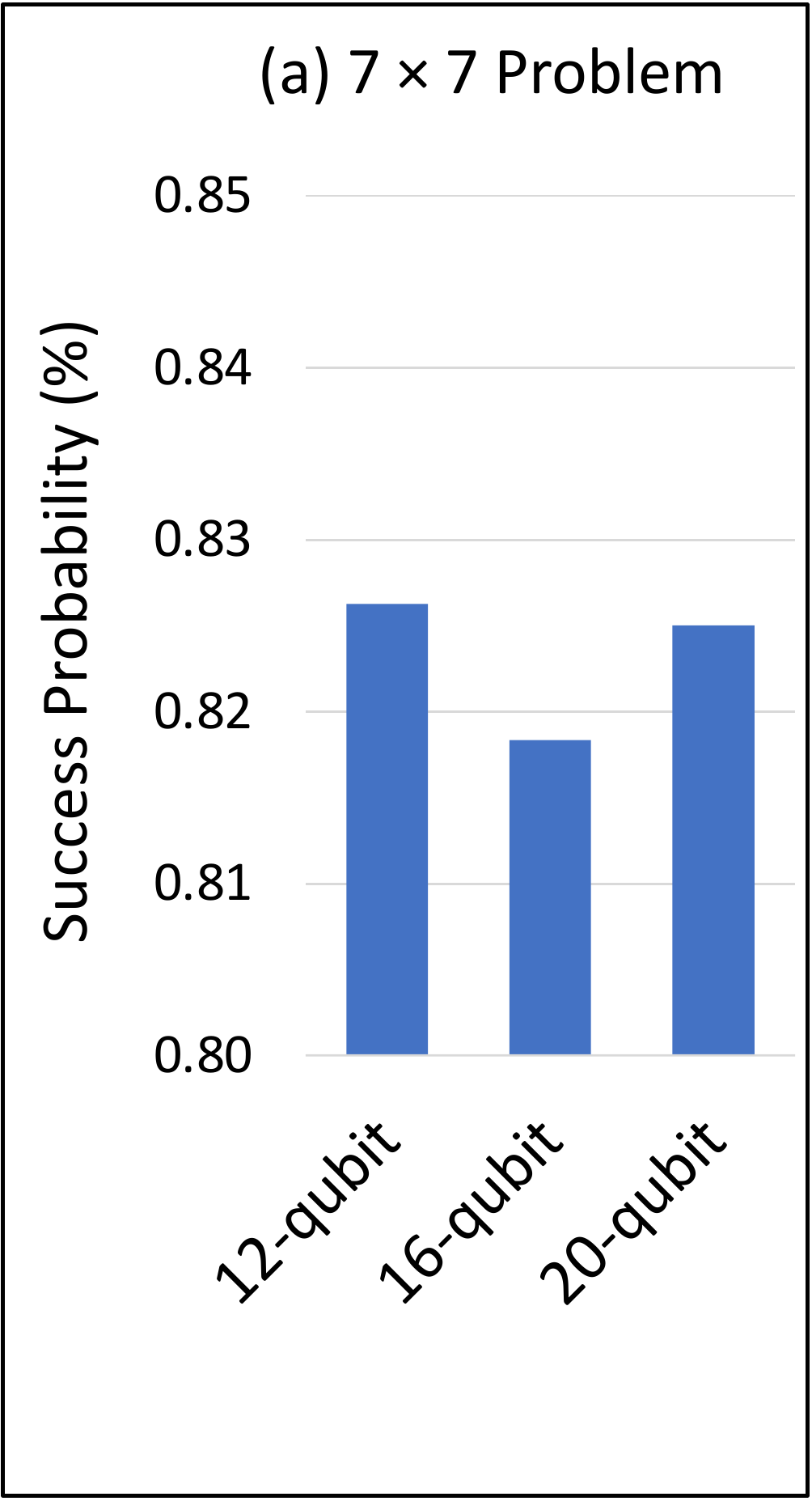}
    \includegraphics[scale=.272]{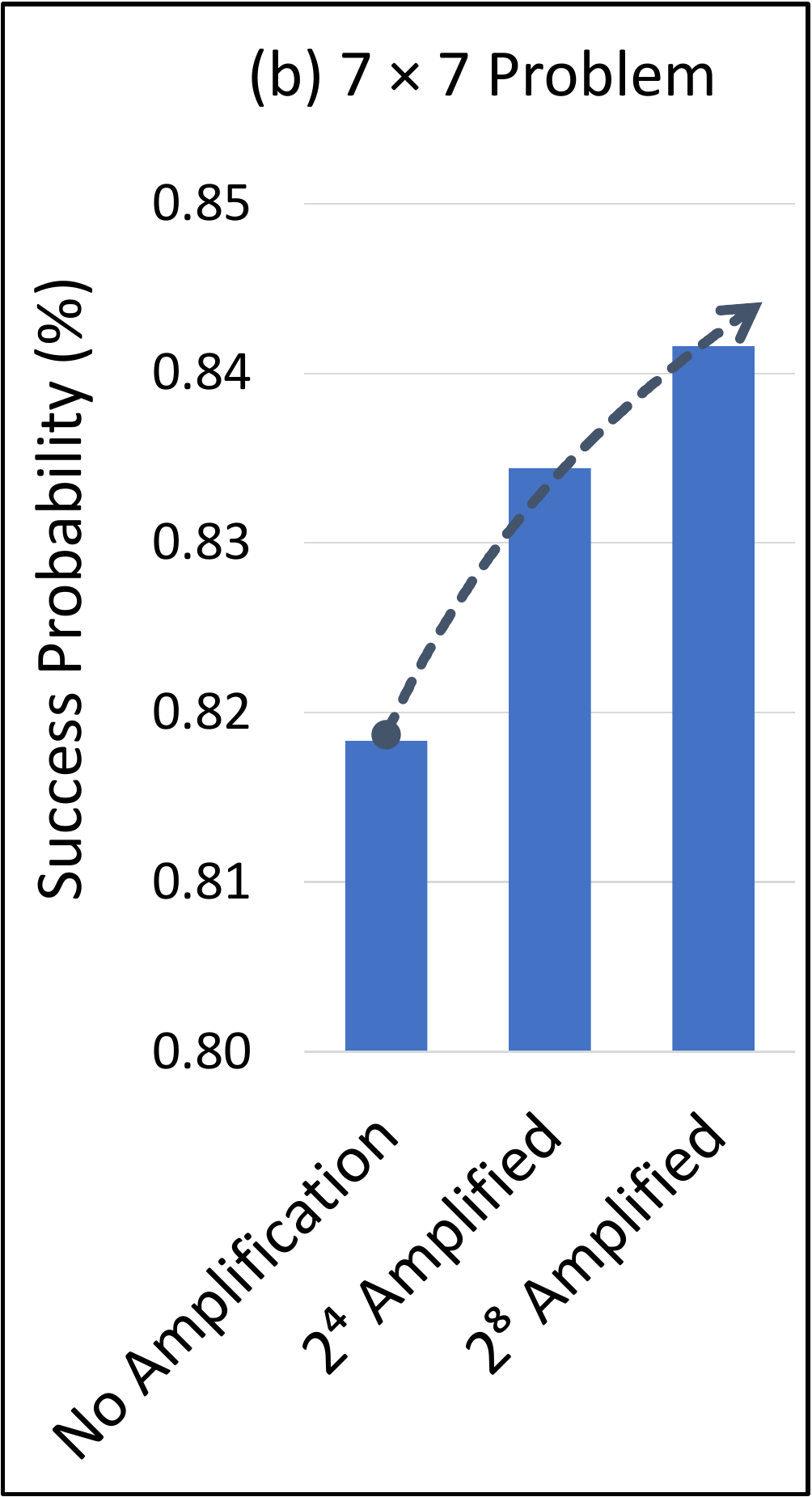}
    \includegraphics[scale=.272]{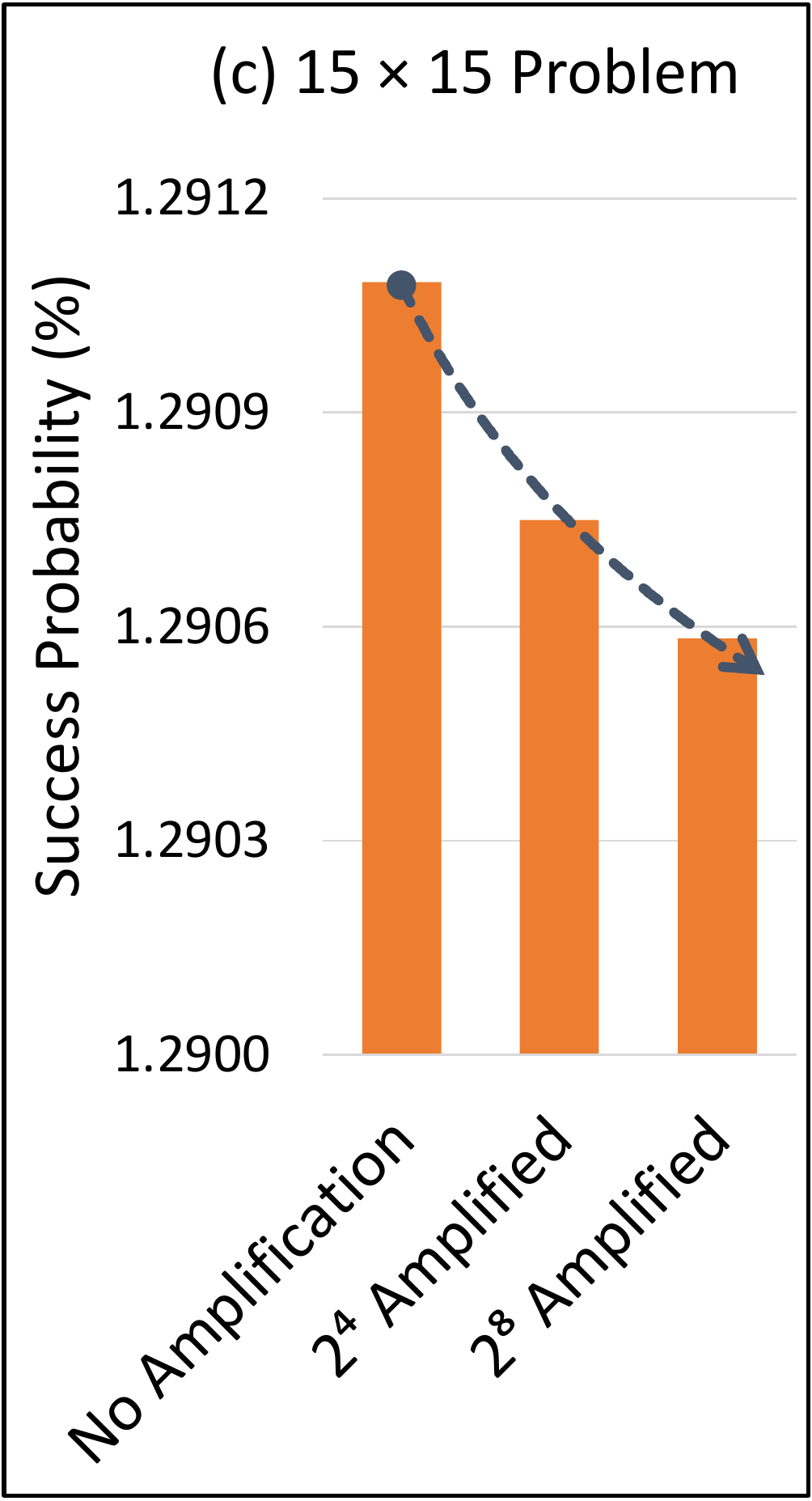}
    \caption{Shows success probability of obtaining the desired state on $7 \times 7$ and $15 \times 15$ problem sizes. Here (a), (b), and (c) correspond to the cases presented in Figs.\,\ref{fig.6}, \ref{fig.7}, and \ref{fig.8}, respectively. The arrow-line in (b) and (c) indicates the steady improvements of success probability towards their respective analytical values with the increase of eigenvalue amplification.}  
    \label{fig.9}
\end{figure}

\textsl{Success Probability.} Analytically, the success probability (SP) of the measurement is determined by the eigenvalue distribution and their levels of accuracy. As can be seen in the state before the measurement, i.e., $|0\rangle\otimes\sum_{j=1}^{2^n - 1} \beta_j  |u_j\rangle\left(\sqrt{1-C^2/\lambda_j^2}|0\rangle + C/\lambda_j|1\rangle \right)$ (C being a normalizing constant) \cite{HHL2009}, the SP is determined by the summation of the squares of reciprocals of eigenvalues. So the values of SP using the truncated (i.e., integer) eigenvalues of $3\times 3$ and $7\times 7$ problems are $1.367\%$ and $1.337\%$, respectively (assuming $C = 1$). However, on the simulation side, we compute the SP by dividing the number of trials with correct output by the total number of repeated trials and then multiplying the factor by 100 \cite{Qi2022}. When no eigenvalue amplification is used, compared to the analytical SP of $1.367\%$ on a $3\times 3$ problem, we computed an SP of $1.103\%$, which is very close to the number $1.120\%$ reported by Wang et al. \cite{Wang2020a}. On the $7\times 7$ problem, as shown in Fig.\,\ref{fig.9}\,(a), the SP appears to vary between $0.818\%$ and $0.826\%$, but without showing any steady movement toward its analytical value $1.337\%$ as more accurate $|\omega_j\rangle$'s were used by increasing the number of qubits in reg.\,A. This suggests that the SP is more sensitive to the other dominant source of error, that involving the truncation of eigenvalues used in phase estimation. Therefore, controlling such error requires using eigenvalue amplification. Figs.\,\ref{fig.9}\,(b) and (c) show the SP on $7\times 7$ and $15\times 15$ problems plotted with different level of amplification. As expected, both figures confirm the steady improvements of the SP rightly proceeding to their analytical values $1.154\%$ and $1.122\%$ (those calculated using the exact eigenvalues), respectively, with higher levels of amplification. Though we have no doubt that a higher amplification factor ($f>8$) would further improve the SP approaching it to its respective analytical value, we are unable to fully characterize the reason for two different variation trends of SP shown by the dotted-arrow in Figs.\,\ref{fig.9}\,(b) and (c). A potential reason could be due to the fact that for the $15\times 15$ problem, we do not have a polynomially greater number of  trials than that of the $7\times 7$ problem as required by the relation of $\kappa$ with $N$ (the size of the discretized matrix $A$). Also, as $\kappa$ grows, matrix $A$ becomes more and more difficult to invert, and the solutions become less stable \cite{HHL2009}. Furthermore, the basic error of the solutions caused by the central-difference approximation is related to the condition number as $\kappa = O(\epsilon^{-2\alpha})$ ($\epsilon$ being error and $\alpha$ being a smoothness parameter), and therefore, an additive preconditioner \cite{Pan2010} may be used to reduce $\kappa$.

\textsl{Summarizing the Input and Output.} In Table 1, we present all the problems we discussed so far, along with each input state $|b\rangle$ and Poisson solution $|v\rangle$. Note that the relative errors shown in Table 1 gradually increase with the problem size. This may be explained by the fact that even a small inaccuracy in the encoded eigenvalues would cause the accumulation of a larger amount of error due to the extra controlled-$U$ operations required for larger problems. Therefore, an optimum solution of a larger problem would require using even higher factors of amplification. Also, for all of our simulations, even though we use angular coefficients encoded to a fixed number of qubits (16), encoding them to a higher number of qubits would certainly improve the accuracy of the solution.

\begin{figure}
    \centering
    \includegraphics[scale=.272]{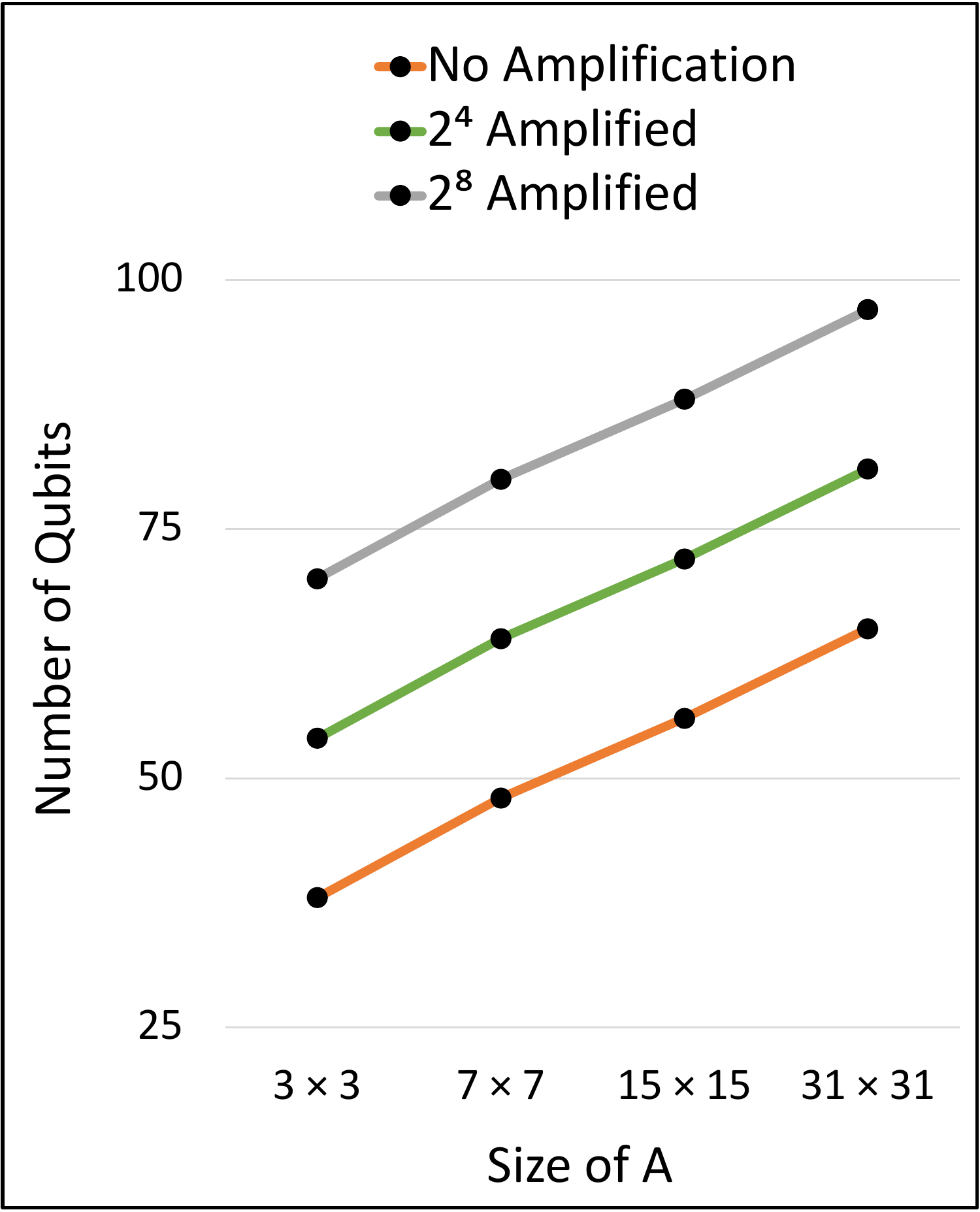}
    \includegraphics[scale=.272]{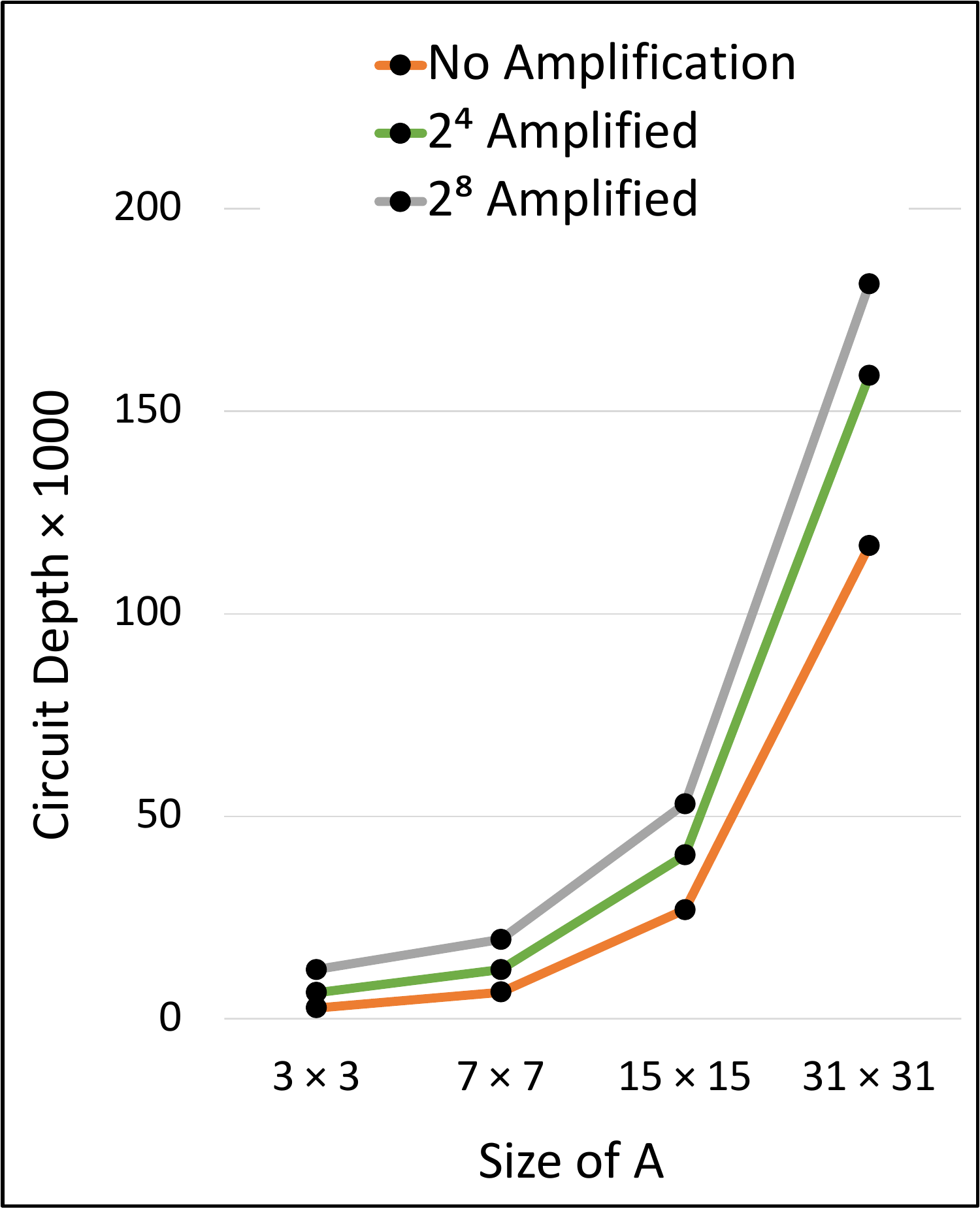}
    \caption{An optimum number of qubits is used for constructing the circuit representing the Poisson solver algorithm. (Left) Shows how the  resources, that is, the number of qubits, scales with the size of the problem. The scaling is shown with varying levels of amplification. (Right) Shows the circuit depth with the size of the problem.}
    \label{fig.10}
\end{figure}

\textsl{Algorithm Scaling and Further Improvement Direction.}
Compared to Cao's algorithm \cite{Cao2013}, Wang's method reduces the cost of the problem by one order, from $O(m^4)$ to $O(m^3)$, by performing the controlled rotation of HHL using the arc cotangent function, meaning the rotation angles are prepared directly from the eigenvalues instead of their reciprocals. Our algorithm, following Wang's approach, not only ensures better accuracy of solutions that are lacking in the existing approaches \cite{HHL2009, Cao2013, Wang2020a} but also successfully demonstrates the scaling of the problem to larger matrices. Within our implementation, our codebase dynamically generates optimized circuits for any given size of the problem. During runtime, we recorded the total number of qubits used and the circuit depth on the basis of elementary gates after the circuit decomposition. As shown in Fig.\,\ref{fig.10}, though the circuit depth grows exponentially as the problem size increases, the number of qubits scales linearly, which is encouraging. This is because, for a simulator or quantum hardware, a critical limiting factor is the total number of qubits, but not the circuit depth. 

However, one may also point out that the exponential increase of the circuit depth may require a longer coherent time, which indeed is still a challenge to increase from the technological development point of view. The circuit depth issue and the overall scaling can be further optimized by: (1) Optimally mapping the logical to physical qubits when compiling quantum circuits onto hardware with restricted connectivity by trading off circuit depth and gate count \cite{Li2019}. We have already implemented this and discuss more about it later in the experimental section; (2) Combining our algorithm structure with an iterative method \cite{Saito2021} would further optimize qubit usage, especially for the eigenvalue expression, while improving the computational speedup by requiring fewer repeated measurements. In fact, our algorithm is well suited for coupling with an iterative solution process, which would ensure even higher accuracy in results; and (3)
Adapting a circuit knitting technique \cite{Eddins2022, Bravyi2016, Peng2020, Tang2021}, which allows partitioning of large quantum circuits into subcircuits that fit on smaller devices, and then knitting the results back together using a classical computer. Although there is some overhead associated with the knitting process, it would open a path to explore massive problems, including multidimensional ones. In our current implementation, due to the full circuit being processed in a single quantum processor, the section of the workflow is relatively slow, especially on large problems. Circuit knitting would require locating processing bottlenecks through profiling and accordingly distributing the tasks on multiple quantum processing units (QPUs), ensuring the tasks' parallelism with load-balancing, which would result in the speeding up of the whole computation. In fact, this is the path IBM takes in realizing their near-term hardware development by combining multiple QPUs \cite{Bravyi2022, Tham2022, Piveteau2023} through circuit knitting techniques.

\renewcommand{\arraystretch}{2.0}
\begin{table*}
\centering
\begin{tabular}{p{1.35cm}|p{5.5cm}|p{7.0cm}|p{1.5cm} }
%  \hline
%  \multicolumn{3}{c}{Country List} \\
 \hline
 Size of $A$ & \centering{Input States, $|b\rangle$} & \centering{MPS Simulated Poisson Solution, $|v\rangle$} & Relative Error~(\%) \\
 \hline\hline 
 $3 \times 3$   
 & $\frac{1}{\sqrt{2}}|01\rangle + \frac{1}{2}(|10\rangle + |11\rangle)$    
 & $0.553|01\rangle + 0.673|10\rangle + 0.490|11\rangle$ 
 & $0.0899$ \\
 \hline
 $7 \times 7$
 & $\frac{1}{4}(|001\rangle + |010\rangle + |011\rangle + |100\rangle) + \frac{1}{2}(|101\rangle + |110\rangle + |111\rangle) $   
 & $0.184|001\rangle + 0.323|010\rangle + 0.418|011\rangle + 0.473|100\rangle + 0.484|101\rangle + 0.409|110\rangle + 0.248|111\rangle$ 
 & $0.1839$ \\
 \hline
$15 \times 15$      
 & $\frac{1}{4}(|0001\rangle + |0010\rangle + |0011\rangle + |0100\rangle + |0101\rangle + |0110\rangle + |0111\rangle + |1000\rangle + |1001\rangle + |1010\rangle + |1011\rangle + |1100\rangle) + \frac{1}{2}|1101\rangle + 0.000|1110\rangle + 
 0.000|1111\rangle$ 
  & $0.080|0001\rangle + 0.152|0010\rangle + 0.209|0011\rangle + 0.258|0100\rangle + 0.297|0101\rangle + 0.322|0110\rangle + 0.340|0111\rangle + 0.344|1000\rangle + 
 0.337|1001\rangle + 0.319|1010\rangle + 0.292|1011\rangle + 0.257|1100\rangle + 0.208|1101\rangle \ \ + \ \ 0.139|1110\rangle \ \ +  \ \ 0.070|1111\rangle$ 
 & $0.5825$ \\ 
 \hline
\end{tabular}
\caption{Shows one-dimensional Poisson equation, i.e., $A|v\rangle=|b\rangle$'s inputs and solution states, and relative errors in the solutions for different sizes of problems.}
\label{table-1}
\end{table*}

If we extend this to $d$ dimensions, the main difference would be the Hamiltonian simulation for $e^{iA^{(d)}t}$, which can be parallelized across $d$ (see Eq.\,\ref{eq.4c}). Therefore, for the multidimensional case, the complexity of our algorithm still grows linearly. This is encouraging as the cost of any classical algorithm solving the $d$-dimensional Poisson equation grows exponentially with $d$. The linear cost of the quantum algorithm makes it ideal for experiments solving the $d$-dimensional Poisson equation on near-term quantum hardware and achieving exponential speedup in terms of $d$.

\section{Circuit Demonstration on Quantum Hardware}
Qiskit allows for easy circuit optimization and the running of circuits on IBM's quantum hardware \cite{Qiskit}. Their \textsl{ibmq{\textunderscore}manila} and \textsl{ibmq{\textunderscore}brooklyn} systems containing 5 and 65 qubits, respectively, \cite{IBMQ} are used to run our circuit experiments. These systems support only the \textsl{CNOT}, $I$, $R_z$, $\sqrt{X}$, and $X$ gates, so any other gates used must be compiled down to these basic components; for example,  the MCMT operation is compiled to \textsl{CNOT} gates. The circuit transformation is performed using Qiskit's transpiler \cite{Younis2022} with samplings that ensure a minimum depth of the optimized circuit.

One crucial part of experimenting with circuits on physical hardware is finding the optimal mapping of virtual qubits to physical qubits on the hardware \cite{Li2019, Zulehner2019, Wille2019, Murali2019, Burgholzer2022}. Qiskit does this automatically via stochastic mappings of virtual to physical qubits and offers different levels of transpilation for circuit optimization. We experimented with multiple levels of optimization, conducting stochastic searches of mappings in an effort to find an optimal mapping for our circuit. Our final circuit was scholastically sampled over 1500 times to find such an optimal mapping. However, as we will discuss, the accuracy of our experiments was ultimately limited by the accumulated error of the large number of \textsl{CNOT} gates required in our circuit.

\subsection{Measurement Error Mitigation on $|b\rangle$}
The current state of quantum hardware presents many challenges, particularly the short coherence time and accumulation of noise in experiments \cite{Corcoles2019}. In addition, on physical devices like IBM's \textsl{ibmq{\textunderscore}manila} or \textsl{ibmq{\textunderscore}brooklyn}, different pairs of qubits have different \textsl{CNOT} error rates, which also affects the ultimate accuracy of the system as many qubits are directly entangled with other qubits using \textsl{CNOT} gates in the course of an experiment \cite{IBMQ}. Therefore, it makes sense to first set up a small system with a limited number of \textsl{CNOT} gates and to experiment on that. 

Additionally, there are two more purposes for this experiment: (1) Setting up a test model with the exact input state used in the full circuit for the $3\times 3$ problem (corresponding to Figs.\,\ref{fig.4} and \ref{fig.5}), from which we get an estimated error related to the measurement part of the algorithm; and (2) Determining how much of that measurement error may be mitigated through the existing model and how the error associated with the relatively small number of CNOT gates affects the overall result.

\begin{figure}[t]
    \centering
    \includegraphics[scale=.160]{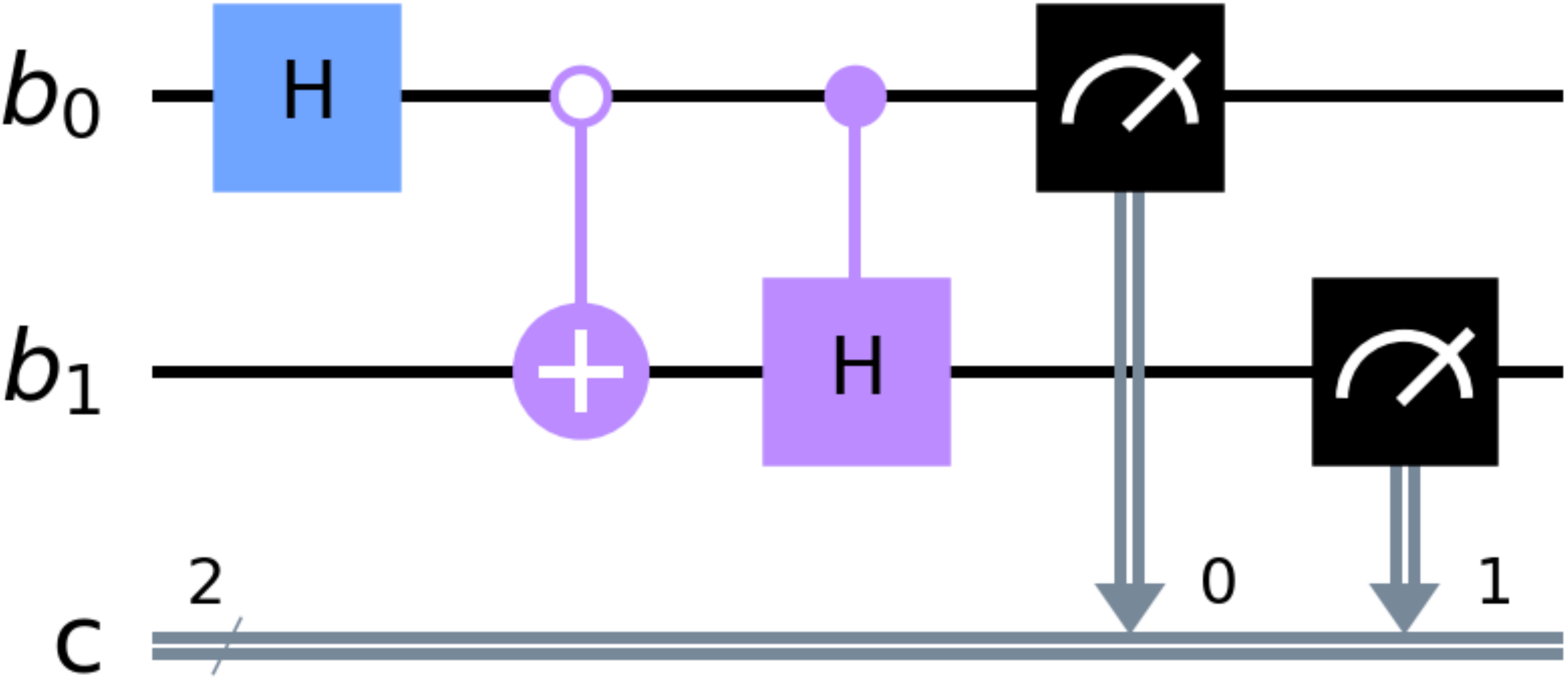}
    \includegraphics[scale=.27]{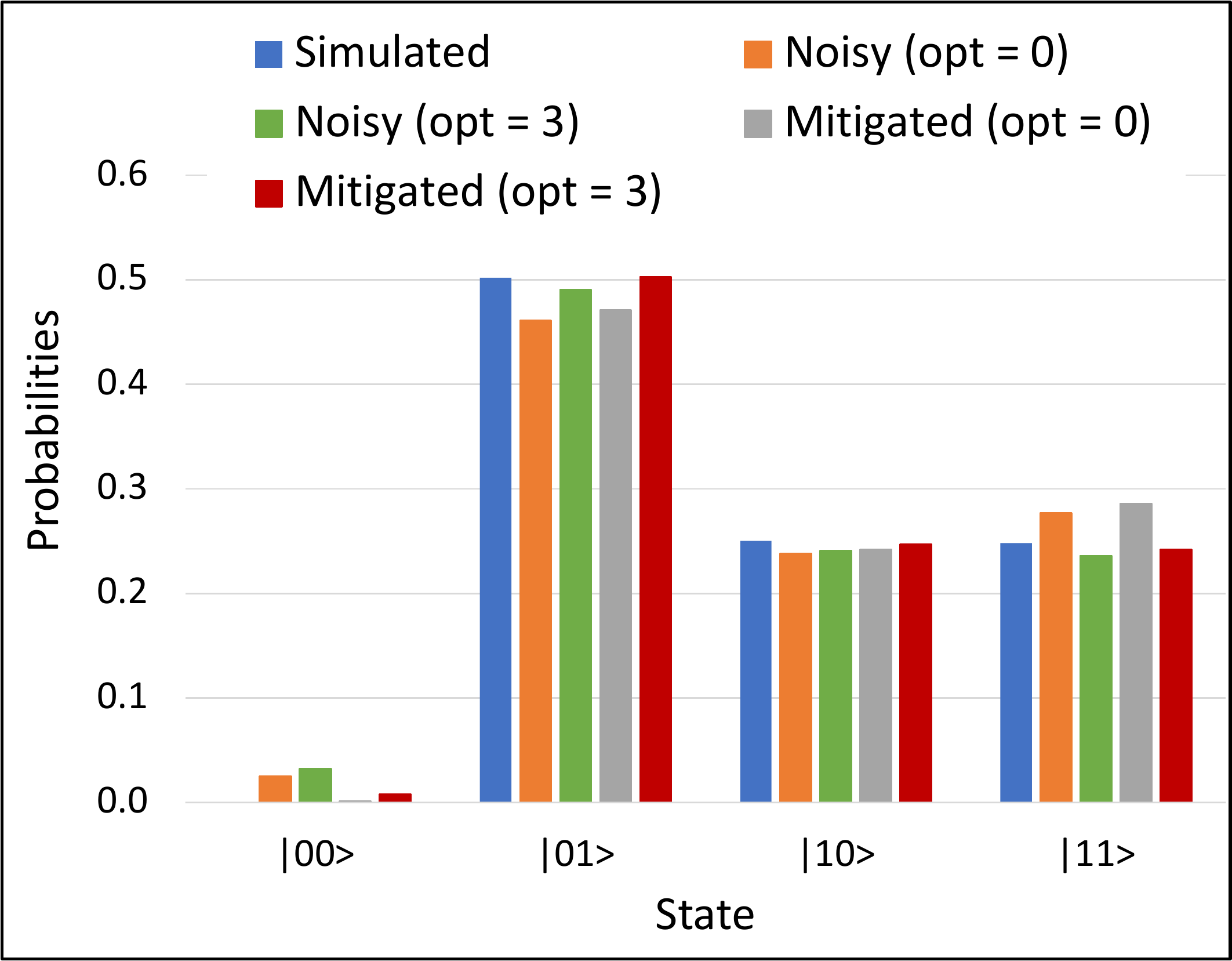}
    \includegraphics[scale=.27]{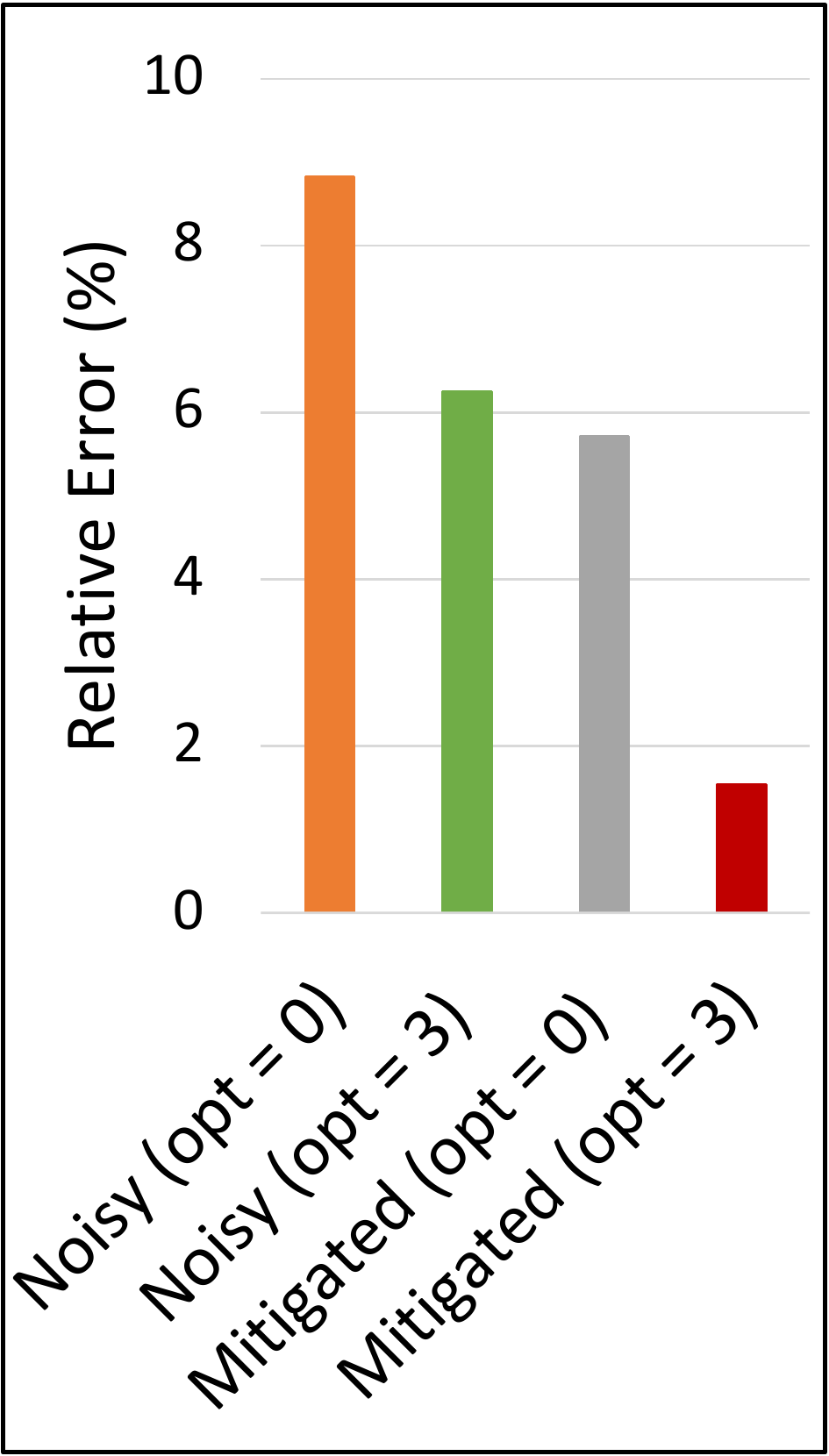}
    \caption{Measurement error mitigation of the simplified circuit built out of $3\times 3$ problem. (Top) Quantum circuit built using the input state $|b\rangle = \frac{1}{\sqrt{2}}|01\rangle + \frac{1}{2}(|10\rangle + |11\rangle)$ of the $3 \times 3$ problem. (Bottom-left) The MPS-simulated result is compared with the experimental result from the noisy IBM's \textsl{ibmq{\textunderscore}manila} device and after mitigating the error on the measurement qubits. Results are shown with the optimization levels $0$ and $3$ of transpiler. (Bottom-right) The relative error in the experimental results with respect to MPS-simulated result.} 
    \label{fig.11}
\end{figure}

Based on the available options for experimentation, we first investigate errors using a simple noise model generated from the properties of real device \textsl{ibmq{\textunderscore}manila} from the IBM Quantum \cite{IBMQ} and mitigate those errors on the measurement qubits \cite{Ferris2022, Funcke2022, Alexandrou2021, Acampora2021}. 
To estimate the amount of error in our actual circuit, it was enough to use a test circuit involving only the input/output state $|b\rangle$ (those acting on reg.\,B) where we do the measurement. A diagram of the circuit is shown in the top panel of Fig.\,\ref{fig.11}. Following the circuit transformation through the transpiler for \textsl{ibmq{\textunderscore}manila}, the circuit decomposes to a number of basis gates that includes $2$ \textsl{CNOT} gates and a few single qubit gates. We experiment with the circuit on \textsl{ibmq{\textunderscore}manila} with and without mitigating errors on both measurement qubits and compare those results with the MPS-simulated result. The results with two optimization levels $0$ and $3$ are shown in the bottom-left panel of Fig.\,\ref{fig.11}. While there are some noticeable differences in probability for some states, the overall result appears to improve with the error mitigation and for higher levels of transpiler optimization. This is clearly evident in the bottom-right panel of the figure, where it shows the relative error with respect to the simulation. It confirms that to reduce the error in the experiment significantly, it is not enough just to tune the optimization levels; error mitigating is also essential on the NISQ hardware. 

We want to mention that while this experiment does not completely represent the full circuit of the Poisson equation solver, we believe that it offers us a projection as to what one could expect if NISQ or near-term hardware could support the experiment of the full circuit. Our experimental results of this test system project a significant reduction of the relative error in the measurement part of the circuit, hence indicating the possibility of mitigating a similar magnitude of error on the full system.

\subsection{Experimenting with the $3\times 3$ Problem on Quantum Hardware}
One measure of the fidelity of quantum systems is in terms of their \textsl{CNOT} error rates, that is, the accuracy of individually entangled bits when performing a two-qubit \textsl{CNOT} gate \cite{Calderon-Vargas2017, Chow2012}. The average \textsl{CNOT} error on the \textsl{ibmq{\textunderscore}brooklyn} system is $8.094\text{e-}2$; in other words, they have an accuracy of about $0.92$. Thus, we can estimate the overall accuracy of the experiment based on the final number of \textsl{CNOT} gates transpiled from the more abstract circuit, approximated by $0.92^c$ where $c$ is the final number of \textsl{CNOT} gates after transpilation. Ultimately, every Toffoli gate, as well as more complex gates such as MCMT, are transpiled into many \textsl{CNOT}s. After a series of transformations using different levels of transpiler optimization, our best circuit for the $3 \times 3$ problem required roughly 5.5k \textsl{CNOT} gates. Unfortunately, this number is quite large given the experimental fidelity of current NISQ devices. As a result, the accumulated errors of the \textsl{CNOT}-gates result in the washing out of the experimental accuracy, which contributes to the artifact of a nonzero contribution for the $|00\rangle$ state (see the figure in Refs.\,\cite{Robson2022, Robson2022b}). In general, experimenting with the circuit on different IBM hardware would end up with similar results, as the best \textsl{CNOT} accuracy on any system is less than 0.99. Therefore, the \textsl{CNOT} error rate appears to be the most dominant bottleneck in realizing the algorithm on NISQ hardware. This experiment helped us pinpoint this key limiting factor of the NISQ device.
In spite of the instrumental difficulties involving the \textsl{CNOT} errors, for the first time, we showed that such a full circuit can easily be mapped (logical to physical gates) and experimented on the existing quantum hardware.

\section{Conclusions}
We have successfully demonstrated several crucial improvements and optimizations essential for scaling the Poisson Solver to larger problem sizes within a hybrid algorithm. By identifying two major sources of error accumulation in the algorithm, one in the phase estimation involving truncating eigenvalues and the other related to the accuracy of the rotation angular coefficients, we were able to build a circuit implementation that was dynamically tunable with respect to those two sources of inaccuracy. Adding accuracy to the eigenvalues through eigenvalue amplification yielded the best improvements and proved to be necessary when expanding to larger, unsolved problem sizes. Not only did we perform more accurate computations with these amplified eigenvalues, but we also were able to achieve a higher success probability on every single circuit than previously possible with truncated eigenvalues. We presented results on significantly larger problem sizes than previous works, as well as improved accuracy on existing problem sizes. These accuracy improvements also translated to the larger problems we demonstrated. 

Clearly, our algorithm represents an advancement in accuracy and usability, and more closely represents what will be put into real-world applications of this theory in the near future. Scalability is a critical step towards breaking the curse of dimensionality that currently plagues solving the Poisson equation, and our multi-level optimized circuit alleviates many of the pressures holding this technology back by dynamically controlling the problem size and register size of crucial segments of the algorithm.

While we were successful in demonstrating our advancements to the Quantum Poisson Solver on a simulator, current quantum hardware proved to be too error-prone to provide accurate results \cite{Johnstun2021, Corcoles2019, Baum2021}. In spite of that, we were able to demonstrate the improvements in the experimental result on IBM's \textsl{ibmq{\textunderscore}manila} device by mitigating error on the measurement qubits on a simplified circuit built out of an exact input/output state of a $3\times 3$ problem and including a small number of \textsl{CNOT} gates. Ultimately, the accumulated error of the large number of \textsl{CNOT} gates required in the full circuit for the Poisson solver, in conjunction with the number of qubits necessary for larger problems, was the limiting factor in our exploration. However, this work has laid the foundation for advanced algorithms that will become usable in the near future as hardware improvements continue. As we see the arrival of more accurate systems with lower \textsl{CNOT} error rates, our algorithm will become usable in larger and more practical problems.  

We have also discussed a vision of how the problem size can be further extended while managing the circuit width and depth at a level suitable to the current technology. In this regard, we have prescribed multilevel solutions, including combining an iterative framework as proposed by Saito et~al.\,\cite{Saito2021}, in order to ensure even higher accuracy in results with fewer repeated shots while requiring an optimum number of qubits. Encouraged by the industry's near-term hardware development roadmap (such as IBM's upcoming quantum-centric supercomputing hardware \cite{Bravyi2022}, for example), we proposed partitioning large circuits through circuit knitting techniques and then running the subcircuits on multiple QPUs in parallel. This would allow us to explore significantly larger problems, including multidimensional ones, with greater computational speed-up.

\begin{acknowledgments}
This research was supported in part by the Notre Dame Center for Research Computing through Notre Dame Research. The authors admire Scott Hampton's comments on the manuscript. The authors also appreciate access to IBM Quantum Hardware through IBM Quantum Network. This work was supported in part by the Department of Energy, Office of Science, Advanced Scientific Computing Research program. This research used resources from the Oak Ridge Leadership Computing Facility at the Oak Ridge National Laboratory, which is supported by the Office of Science of the U.S. Department of Energy under Contract No. DE-AC05-00OR22725. \\
\textsl{Notice}: This manuscript has been authored by UT-Battelle, LLC under Contract No. DE-AC05-00OR22725 with the U.S. Department of Energy.  The publisher, by accepting the article for publication, acknowledges that the U.S. Government retains a non-exclusive, paid-up, irrevocable, world-wide license to publish or reproduce the published form of the manuscript, or allow others to do so, for U.S. Government purposes. The DOE will provide public access to these results in accordance with the DOE Public Access Plan (http://energy.gov/downloads/doe-public-access-plan). 
\end{acknowledgments}

% \nocite{*}

\bibliography{main}% Produces the bibliography via BibTeX.

\end{document}